\PassOptionsToPackage{pdfpagelabels=false}{hyperref} 
\documentclass[fleqn,usenatbib]{mnras}
\usepackage{newtxtext,newtxmath}

\usepackage[utf8]{inputenc} 
\usepackage[T1]{fontenc}    
\usepackage{hyperref}       
\usepackage{url}            

\usepackage[table]{xcolor}

\usepackage{graphicx}	
\usepackage{amsmath}	
\usepackage{bm}

\usepackage{breakcites}
\usepackage{url} 
\usepackage{float}
\usepackage{subfigure}
\usepackage{color}
\usepackage{lineno} 
\usepackage{enumitem}  
\usepackage{chngcntr}
\usepackage{cancel}
\usepackage{threeparttable}
\usepackage{tikz}
\usepackage{academicons}

\usepackage{pifont}
\newcommand{\eqb}{\begin{eqnarray}}
\newcommand{\eqe}{\end{eqnarray}}
\newcommand{\bi}{\begin{itemize}}
\newcommand{\ei}{\end{itemize}}
\newcommand{\sect}[1]{Sec.~\ref{sec:#1}}
 
\newcommand{\tab}[1]{Table~\ref{tab:#1}}

\newcommand{\fermi}{\emph{Fermi}}
\definecolor{cornflowerblue}{rgb}{0.39, 0.58, 0.93}
\definecolor{darkspringgreen}{rgb}{0.09, 0.45, 0.27}
\definecolor{turquoise}{rgb}{0.03, 0.91, 0.87}
\definecolor{coral}{rgb}{0.97, 0.51, 0.47}
\definecolor{indian}{rgb}{1.0, 0.22, 0.0}
\definecolor{darkorchid}{rgb}{0.6, 0.2, 0.8}
\definecolor{lime}{rgb}{0.8, 1.0, 0.0}

\definecolor{lime}{HTML}{A6CE39}
\DeclareRobustCommand{\orcidicon}{%
	\begin{tikzpicture}
	\draw[lime, fill=lime] (0,0) 
	circle [radius=0.16] 
	node[white] {{\fontfamily{qag}\selectfont \tiny ID}};
	\draw[white, fill=white] (-0.0625,0.095) 
	circle [radius=0.007];
	\end{tikzpicture}
	\hspace{-2mm}
}

\foreach \x in {A, ..., Z}{%
	\expandafter\xdef\csname orcid\x\endcsname{\noexpand\href{https://orcid.org/\csname orcidauthor\x\endcsname}{\noexpand\orcidicon}}
}

\newcommand{\orcid}[1]{\href{https://orcid.org/#1}{\textcolor[HTML]{A6CE39}{\orcidicon}}}
\hypersetup{colorlinks,linkcolor={teal},citecolor={teal},urlcolor={teal}}  

\title[Magnetic baryon-loaded jets]{Hadronic signatures from magnetically dominated baryon-loaded AGN jets}
\author[Petropoulou, Psarras, Giannios]{
Maria Petropoulou\orcid{0000-0001-6640-0179}$^{1}$, Filippos Psarras$^{1}$, Dimitrios Giannios$^2$
\\
$^{1}$Department of Physics, National and Kapodistrian University of Athens, University Campus Zografos, GR 15783, Greece\\
$^2$Department of Physics, Purdue University, 525 Northwestern Avenue, West Lafayette, IN 47907, USA
}

\date{Accepted XXX. Received YYY; in original form ZZZ}

\pubyear{2022}

\begin{document}

\label{firstpage}
\pagerange{\pageref{firstpage}--\pageref{lastpage}}
\maketitle

\begin{abstract}
Blazars are a rare class of active galactic nuclei (AGN) with relativistic jets pointing towards the observer. Jets are thought to be launched as Poynting-flux dominated outflows that accelerate to relativistic speeds at the expense of the available magnetic energy. In this work, we consider electron-proton jets and assume that particles are energized via magnetic reconnection in parts of the jet where the magnetization is still high ($\sigma \ge 1$). The magnetization and bulk Lorentz factor  $\Gamma$ are related to the available jet energy per baryon as $\mu =\Gamma(1+\sigma)$. We adopt an observationally motivated relation between $\Gamma$ and the mass accretion rate into the black hole $\dot{m}$, which also controls the luminosity of external radiation fields. We numerically compute the photon and neutrino jet emission as a function of $\mu$ and $\sigma$. We find that the blazar SED is produced by synchrotron and inverse Compton radiation of accelerated electrons, while the emission of hadronic-related processes is subdominant except for the highest magnetization considered. We show that low-luminosity blazars ($L_{\gamma} \lesssim 10^{45}$~erg s$^{-1}$) are associated with less powerful, slower jets with higher magnetizations in the jet dissipation region. Their broadband photon spectra resemble those of BL Lac objects, and the expected neutrino luminosity is $L_{\nu+\bar{\nu}}\sim (0.3-1)\,  L_{\gamma}$. High-luminosity blazars ($L_{\gamma} \gg 10^{45}$~erg s$^{-1}$) are associated with more powerful, faster jets with lower magnetizations. Their broadband photon spectra resemble those of flat spectrum radio quasars, and they are expected to be dim neutrino sources with $L_{\nu+\bar{\nu}}\ll L_{\gamma}$. 
\end{abstract}




\section{Introduction}

Blazars are a rare class of active galactic nuclei (AGN) with relativistic jets that are powered by accretion onto a central supermassive black hole \citep{Begelman1984} and are closely aligned to our line of sight \citep{Urry1995}. They are the most powerful persistent astrophysical sources of non-thermal electromagnetic radiation in the Universe, and are promising candidate sources of other cosmic messengers like high-energy neutrinos
\citep[for a recent review, see][]{2022arXiv220203381M}. 

The spectral energy distribution (SED) of blazars, which extends from radio wavelengths to $\gamma$-ray energies, has a characteristic double-hump shape. The blazar SED is dominated by variable non-thermal emission, which is Doppler boosted as it originates in the relativistic jet pointing toward the observer. Depending on the location of the peak energy, $\varepsilon_{\rm pk}$, of the low-energy hump  (in a $\varepsilon L_{\varepsilon}$ versus $\varepsilon$ plot), blazars are divided in three spectral subclasses~\citep[][]{Abdo_2010}: high-synchrotron peaked (HSP) blazars for $\varepsilon_{\rm pk} \ge 4$~eV, intermediate-synchrotron peaked (ISP) blazars for $0.4~{\rm eV} < \varepsilon_{\rm pk} < 4$~eV, and low-synchrotron peaked (LSP) blazars for $\varepsilon_{\rm pk} < 0.4$~eV. Historically, blazars were also divided in flat-spectrum radio quasars (FSRQs) and BL Lac objects (BL Lacs) based on the appearance of their optical spectra. The former class displays strong, broad emission lines, while the latter class has almost featureless optical spectra, showing at most some weak emission lines~\citep[for a recent review, see][]{Padovani_2017}. It is noteworthy that all FSRQs are LSP blazars, with a few exceptions that belong to the ISP subclass.

The observational differences between FSRQs and BL Lacs have been attributed to differences in the accretion flow of the central black hole. In particular, \cite{Ghisellini_2011, Sbarrato_2012} proposed that FSRQs have accretion discs with bolometric luminosity $L_{\rm d} \gtrsim 10^{-2} L_{\rm Edd}$, where $L_{\rm Edd}$ is the Eddington luminosity of the central black hole. \cite{Padovani_2017} also stressed that the dividing line in $L_{\rm d}/L_{\rm Edd}$ is meaningful in a statistical sense, as the blazar's divide may also depend on other factors like the black hole mass and spin. Moreover, a strong correlation between the jet luminosity and the accretion power was found~\citep[see e.g.][]{Ghisellini_2014}, highlighting the jet-disc coupling in blazars.

\cite{GPP_2013} suggested that some blazars with radiatively efficient discs ($L_{\rm d}/L_{\rm Edd}>0.01$) may appear as \textit{masquerading} BL Lacs, simply because the optical emission lines are swamped by the non-thermal jet continuum \citep[see also][]{1998ApJ...506..621G}. Notably, the first astrophysical source to be associated with high-energy neutrinos was the blazar TXS 0506+056 \citep{IceCube2018_2017txs, IceCube2018_2014txs}, a masquerading BL Lac \citep{2019MNRAS.484L.104P}. Since then several works have reported hints  of possible associations between neutrinos detected by IceCube and blazars at different levels of statistical significance \citep[e.g.][]{IceCube10year, Plavin2020, Hovatta2021,  Padovani_2022,Padovani_2022_pks, Buson_2022}. 
If at least some of these associations are true, then blazar jets should contain baryons that are accelerated to relativistic energies.

Radiation models applied to multi-wavelength observations of TXS~0506+056~\citep[e.g.][]{Keivani2018, Cerruti_2019, Gao_2019} and other individual sources of interest, like PKS 0735+178 \citep{2022arXiv220405060S} and PKS 1502+106 \citep{Oikonomou_2021}, are phenomenological with many free parameters, including those describing the accelerated particles, such as spectrum and total energy. These are left free to vary when modelling the SED in order to maximize the neutrino emission of the source (at energies relevant for IceCube). Moreover, external photon fields, which are not directly visible, are often invoked to enhance the production rate of neutrinos~\citep[e.g.][]{Reimer_2019,Rodrigues_2019}. Despite the heterogeneous approaches used for the modelling of individual sources, some general conclusions can still be drawn. First, \textit{hybrid} radiation models, where the SED is attributed to emission of accelerated electrons, and the radiative signatures of protons are not directly visible, are favoured for all blazar subclasses and, in particular, for LSP/ISP sources. Even though the electromagnetic emission of hadronic-related processes in these models is not imprinted on the SED, the proton power can still be constrained when modelling their putative neutrino emission. 
Second, the baryonic loading of jets, usually defined as the ratio of the bolometric luminosity in relativistic protons and the observed $\gamma$-ray luminosity, is $10 \lesssim L_{\rm p}/L_{\gamma} < 10^5$, with higher values found for lower luminosity blazars. Third, the ratio of the all-flavour neutrino luminosity to the $\gamma$-ray luminosity, $Y_{\nu \gamma}=L_{\nu + \bar{\nu}}/L_{\gamma}$, is smaller than unity and has a decreasing trend with increasing $L_{\gamma}$~\citep[see e.g. Fig.~15 in][]{Petropoulou_2020}. These recent results raise the following question: is there a physically motivated scenario for blazar emission that could explain the trends of $Y_{\nu \gamma}$ with $L_{\gamma}$ while constraining the properties of the accelerated particles in the emitting region of jets? 

AGN jets are thought to be launched from the vicinity of a rotating accreting black hole as Poynting-flux dominated plasma outflows~\citep{BZ77}. Ideal magnetohydrodynamic (MHD)  models
of magnetically dominated jets predict that the bulk acceleration of the jet takes place at the expense of its magnetization $\sigma$ (i.e. the ratio of the Poynting flux and the total energy flux of the jet). This means that the jet bulk Lorentz factor $\Gamma$ increases, as $\sigma$ decreases, while the total energy flux of the jet remains constant in the absence of energy dissipation. Such magnetic acceleration is spatially extended and terminates when the jet becomes matter-dominated, i.e. $\sigma < 1$ \citep{Vlahakis_2004, Komissarov_2007, 2011MmSAI..82...95K}. 

Observations of blazars reveal that their jets are characterized by a high radiative efficiency~\citep[e.g.][]{Celotti_2008, Ghisellini_2014}.
If blazar jets were ideal MHD outflows, i.e. there was no mechanism in place for tapping the available jet energy and transferring it to non-thermal particles, then jets would simply not shine. If energy dissipation takes place in regions where the relativistic bulk flow remains Poynting dominated (i.e. $\sigma \gtrsim 1$), then magnetic reconnection\footnote{This is a process that liberates energy stored in magnetic fields during a topological rearrangement of the field lines. The energy is then transferred to the plasma, both via heating and acceleration of particles.} is a more promising mechanism for particle energization than relativistic shocks~\citep{Sironi_2015}; for a recent review, see also \cite{2020NewAR..8901543M}. Both two-dimensional (2D) and three-dimensional (3D) kinetic simulations of reconnection in the relativistic regime ($\sigma \ge 1$) have shown that particles are efficiently accelerated into power-law distributions with slopes depending on $\sigma$~\citep[e.g.][]{Sironi_2014, Guo_2014, Werner_2016}. More specifically, hard power laws with $p\sim 1.5$ are found for $\sigma \gg 10$, while $p\sim 2.5-3$ is found for $\sigma\sim 1-3$. Recent 3D simulations of reconnection find a weaker dependence of the slope on plasma magnetization, but with the same overall trend \citep{Zhang_2021}. 
This unique dependence of the particle distribution shape on $\sigma$ is bound to have an impact on the observed jet emission~\citep[see e.g.][]{Petropoulou_2016, Christie_2019}. Meanwhile, the fraction of energy transferred to non-thermal particles does not strongly depend on $\sigma$, and can reach 50 per cent in pair plasmas or 25 per cent in electron-proton plasmas \citep{Sironi_2015}.  

Recently, \citet{2021MNRAS.501.4092R} proposed a fairly simple idea that could account for the observed spectra of blazars. According to this, all jets, regardless of their power, are launched with similar energy per baryon, $\mu$. Using an observationally driven correlation between the accretion rate and the jet Lorentz factor, $\dot{m} \propto \Gamma^s$,  \citet{2021MNRAS.501.4092R} proposed that FSRQs have faster more powerful jets, with moderate magnetization in their emission region, which results in steep particle energy distributions. BL Lac objects, on the other hand, are associated with less powerful and slower jets, which retain higher magnetizations in the region where particle energization occurs. The appealing aspect of their model is that all physical quantities can be traced back to two fundamental parameters of the jet, namely $\mu$ and $\sigma$. In this paper we expand upon the work of \citet{2021MNRAS.501.4092R} by considering the radiative signatures of relativistic protons accelerated by magnetic reconnection in blazar jets. Contrary to other works where the properties of the relativistic particles, such as their injection spectrum and energy budget, are left free here we use physically motivated values that are connected to the plasma properties of the blazar jet. 

This paper is structured as follows. In \sect{model} we outline the model and the numerical code used. In \sect{results} we present our results on the expected electromagnetic and high-energy neutrino emission. We discuss our findings in \sect{discussion} and conclude in \sect{conclusion}.

\section{Model}\label{sec:model}
We are interested in computing the photon and neutrino emission from blazar jets using a physically motivated model with as few as possible free parameters. Our goal is to express important physical quantities, such as the jet luminosity, the magnetic field strength, the size of the emitting region, with at least two of the three main parameters: the jet magnetization ($\sigma$), the total energy to rest mass flux ratio of the jet ($\mu$), and the dimensionless accretion rate onto the black hole ($\dot{m}$). 

\subsection{Main parameters}
The total energy flux per unit
rest-mass energy flux, $\mu$, is one of the integrals of motion (i.e. quantities that remain constant along magnetic field lines) in an ideal magnetohydrodynamical (MHD) axisymmetric outflow  \citep{Komissarov_2007,Tchekhovskoy_2009}. For a cold outflow, where the pressure and internal energy is negligible compared to the rest mass energy of the plasma, $\mu$ can be written as
\begin{equation} \label{eq:1}
\mu = \Gamma(1+\sigma)  
\end{equation}
where $\Gamma$ and $\sigma$ are the bulk Lorentz factor of the flow and its magnetization, respectively. The latter is usually defined as
\begin{equation}
    \sigma = \frac{B^{'2}}{4\pi \rho' c^2}
    \label{eq:sigma1}
\end{equation}
where $B'$ and  $\rho'$ are the magnetic field strength and mass density in the jet rest frame respectively; henceforth, primed quantities are used to denote quantities in the jet rest frame. For jets with an electron-proton plasma composition, $\rho' \approx n' m_{\rm p}$ where $n'$ is the comoving number density of cold electrons (or protons). According to Eq.~(\ref{eq:1}), if all jets were launched with the same $\mu$, those with lower magnetizations would be asymptotically faster, and vice versa. Moreover, Eq.~(\ref{eq:1}) expresses energy conservation along the jet. MHD models of axisymmetric stationary outflows show that their bulk Lorentz factor increases with distance from the central engine at the expense of the outflow's electromagnetic energy flux~\citep[e.g.][]{Komissarov_2007}. 

The other main parameter in our model, which also distinguishes FSRQs from BL Lac objects, is the dimensionless accretion rate,
\begin{equation} \label{eq:2}
    \dot{m}=\frac{\dot{M}}{{\dot{M}_{\rm Edd}}}
\end{equation}
where $\dot{M}$ is the accretion rate onto the black hole and $\dot{M}_{{\rm Edd}}$ is the Eddington mass accretion rate. This is defined as
\begin{equation} \label{eq:3}
    \dot{M}_{{\rm Edd}}\equiv \frac{L_{\rm Edd}}{\eta_{\rm d}c^2}
\end{equation}
where $L_{\rm Edd}=1.26 \times 10^{38} \left(M/M_{\odot}\right)$~erg s$^{-1}$ is the Eddington luminosity of a black hole with mass $M$, and ${\eta}_{\rm d}$ is the radiative efficiency of the disc. Accordingly, the bolometric disc luminosity is given by
\begin{equation}
    L_{\rm d} = \eta_{\rm d} \dot{M} c^2 = \dot{m} L_{\rm Edd}.
    \label{eq:Ld}
\end{equation}
The radiative efficiency may vary from $\sim6$ per cent to $\sim 40$ per cent in different accreting regimes. For low enough accretion rates ($\dot{m} \lesssim 0.02$), the disc structure changes and becomes radiatively inefficient \citep[see e.g.][for state transitions in X-ray binaries]{Maccarone_2003}. Here, we adopt $\eta_{\rm d} =0.1$ as a default value for all accretion rates considered ($\dot{m}\sim 10^{-5}-1$), and discuss the effects of a mass accretion-dependent radiative efficiency in Sec.~\ref{sec:discussion}. 

Following \cite{2021MNRAS.501.4092R}, we assume that the jet bulk Lorentz factor and the accretion rate are correlated. In particular, \cite{Lister_2009} 
demonstrated that there is a strong correlation between the apparent jet speed and the apparent radio luminosity using a big sample of AGN that were observed with MOJAVE. \cite{Cohen_2007} also investigated the correlation between the intrinsic jet luminosity and the Lorentz factor by performing Monte Carlo simulations, favouring a positive correlation. Motivated by these results, and assuming a jet-disc connection, we use the following relation between $\dot{m}$ and $\Gamma$,

\begin{equation} \label{eq:5}
\frac{\dot{m}}{\dot{m_{\rm o}}} = \left(\frac{\Gamma}{{\Gamma}_{\rm o}} \right)^s 
\end{equation}
where $s>0$. \cite{2021MNRAS.501.4092R} performed a series of simulations with different values of $s$, showing that their results did not change much for $ 2.4 \le s \le 3.5 $. Based on these results we adopt $s=3$ as our default value. Moreover, \cite{Lister_2019} presented estimates of the distribution of maximum jet speeds using another MOJAVE sample of $409$ radio-loud AGN, and it was found that ${\Gamma}_{\rm max} \approx 50$. Consequently, we set $({\Gamma}_{\rm o}, \dot{m}_{\rm o})  = (40, 1)$ to cover also cases with dimensionless accretion rates up to 2 (for selected values of $\mu$ and $\sigma$). Based on the above, Eq.~(\ref{eq:5}) can be rewritten as
\begin{equation}
    \dot{m} \simeq 1.6 \times 10^{-5} \, \Gamma^3 = 1.6 \times 10^{-5} \frac{\mu^3}{(1+\sigma)^3}, 
    \label{eq:mdot}
\end{equation}
where we used Eq.~(\ref{eq:1}) to obtain the expression on the right hand side. According to this equation, jets launched with the same $\mu$ and low magnetizations are powered by higher accretion rates, which in turn are associated with higher disc luminosities (see Eq.~\ref{eq:Ld}), and vice versa. 

\subsection{External radiation fields}
A crucial parameter in our model is the radiation energy density produced by the emission of the Broad Line Region (BLR). This is believed to be reprocessed radiation from the accretion disc. Assuming that the emission is isotropic in the black hole rest frame with typical photon energy $\epsilon_{\rm BLR}=2$~eV, the integrated BLR energy density is given by
\begin{equation} \label{eq:uBLR-1}
u_{\rm BLR} = \frac{ \eta_{\rm BLR} L_{\rm d}  }{4{\pi}cR_{\rm BLR}^2},
\end{equation}
where the BLR radius is estimated as $R_{\rm BLR} = 10^{17}L_{\rm d,45}^{1/2}$~cm and ${\eta}_{\rm BLR} = 0.1 \, \eta_{\rm BLR,-1}$ is the covering factor \citep{Ghisellini_2008}. Here, we introduced the notation $q_x = q/10^x$ (in cgs units). 

The non-thermal emission region is described as a spherical blob in the comoving frame of the outflow. We further assume that this is located, in all cases, close to the outer edge of the BLR at a distance
\begin{equation}
R_{\rm em} = 0.9 \, R_{\rm BLR} = 2.8\times 10^{17}~ \dot{m}^{1/2} \, L_{\rm Edd, 46}^{1/2} \,  {\rm cm}.
\label{eq:Rem}
\end{equation}
While the location of the $\gamma$-ray emitting region in AGN remains an open issue, our choice for $R_{\rm em}$ is motivated by recent results about TXS 0506+056. \cite{2019MNRAS.484L.104P} showed using $\gamma \gamma$ opacity constraints that the emission region cannot reside well within its BLR, but it should be closer to its outer edge. However, it is still possible that $R_{\rm em} > R_{\rm BLR}$. We discuss in this case how our results would be modified in Sec.~\ref{sec:discussion}. 

In a conical jet with half-opening angle ${\theta}_{\rm j} \sim 1/\Gamma$ we can also relate the comoving radius of the emitting blob with $\Gamma$ or $\dot{m}$ as follows
\begin{equation} \label{eq:Rb}
R'_{\rm b} = R_{\rm em}{\theta}_{\rm j} \simeq 1.1\times 10^{15} \left(\frac{\dot{m}}{1.6 \times 10^{-5}} \right)^{1/6} L_{\rm Edd, 46}^{1/2} \, {\rm cm}
\end{equation}
where we used Eqs.~(\ref{eq:mdot}) and (\ref{eq:Rem}).

Moreover, the energy of BLR photons in the blob comoving frame is boosted as $\epsilon'_{\rm BLR} \approx \Gamma\epsilon_{\rm BLR}$ and the comoving energy density reads $u'_{\rm BLR}  \approx {\Gamma}^2 \left(1+ \beta^2/3\right)u_{\rm BLR}$, where $\beta = \sqrt{1-1/\Gamma^2}$. Using Eqs.~(\ref{eq:mdot}) and (\ref{eq:uBLR-1}) and assuming $\beta \approx 1$, the comoving energy density of external photons can be expressed as function of $\Gamma$ or $\dot{m}$,
\begin{equation}
    u'_{\rm BLR} \simeq \frac{1}{18 \pi}\Gamma^2 = \frac{1}{18 \pi} \left( \frac{\dot{m}}{1.6 \times 10^{-5}}\right)^{2/3} \, {\rm erg} \, {\rm cm}^{-3}.
    \label{eq:uBLR}
\end{equation}
\subsection{Jet power and energy dissipation}
The jet power is ultimately connected to the accretion power \citep[e.g.][]{Celotti_2008,Ghisellini_2014} as
\begin{equation} \label{eq:Lj}
L_{\rm j} = \eta_{\rm j} \dot{M} c^2 = \frac{\eta_{\rm j}}{\eta_{\rm d}} \dot{m}  L_{\rm Edd} 
\end{equation}
where ${\eta}_{\rm_j}$ is the jet production efficiency. As a reference value we use $\eta_{\rm j}=0.9$ even though values as high as $\sim 1.4$ are possible in specific accretion regimes and for maximally spinning black holes \citep[e.g.][]{2011MNRAS.418L..79T}. 

The power of a cold outflow is comprised of two main components, one related to matter (kinetic power, $L_{\rm kin}$) and another one related to the electromagnetic fields (Poynting luminosity, $L_{\rm B}$). The magnetization introduced in Eq.~(\ref{eq:sigma1}) can also be defined as
\begin{equation} \label{eq:sigma2}
\sigma = \frac{L_{\rm B}}{L_{\rm kin}} = \frac{L_{\rm B}}{L_{\rm j} - L_{\rm B}}
\end{equation}

From the equation above we can derive the comoving magnetic field strength
\begin{equation} \label{eq:B}
B' = \left(\frac{4 \sigma L_{\rm j}}{(1+\sigma) R^{'2}_{\rm b}c\beta{\Gamma}^2}\right)^{1/2}.
\end{equation}
Combining Eqs.~(\ref{eq:mdot}), (\ref{eq:Rb}) and (\ref{eq:Lj}) the comoving magnetic field turns out to be a function of $\sigma$ alone, namely
\begin{eqnarray}
    B' & = & \left(\frac{4 \sigma}{(1+\sigma) c\beta} \right)^{1/2} \left( \frac{\eta_{\rm j}}{\eta_{\rm d}}\right)^{1/2} \nonumber \\ 
    & \simeq & 12~\left(\frac{\sigma}{1+\sigma}\right)^{1/2}\left(\frac{\eta_{\rm j}}{0.9}\right)^{1/2} \left(\frac{\eta_{\rm d}}{0.1}\right)^{-1/2}~{\rm G},
\end{eqnarray}
where $\beta \approx 1$ was used when deriving the numerical value. This approximation breaks down for combinations of $\mu$ and $\sigma$ that lead to $\Gamma \sim 1$. For example, for $\mu=50$ and $\sigma=48.9$, 
the magnetic field strength reaches a maximum value of $\sim 48$~G and not $\sim 12$~G as predicted by the approximate expression above. An even stronger magnetic field in the emitting region could be achieved, if this was located closer to the black hole (e.g.  $B'\sim 100$~G for $R_{\rm em} = 0.1\, R_{\rm BLR}$). 

In non-thermal emitting astrophysical outflows there should be a mechanism in place for dissipating energy (carried by the matter in form of kinetic or thermal energy or by the electromagnetic fields) and transferring it into non-thermal radiating particles. Magnetic reconnection is often considered as a primary process for fast energy release and particle energization in magnetically dominated environments \citep[for a recent review, see][]{2020PhPl...27h0501G}. 

The luminosity transferred to non-thermal electrons and protons (as measured in the comoving frame of the blob) can be related to the Poynting jet luminosity as 
\begin{equation} \label{eq:Lpart}
L'_{\rm e} = L'_{\rm p} = f_{\rm rec} \frac{2 L_{\rm B}}{3\beta{\Gamma}^2}
\end{equation}
where $f_{\rm rec}$ is the fraction of the dissipated magnetic energy that is distributed to relativistic particles. Using 2D particle-in-cell simulations, \cite{Sironi_2015} estimated $f_{\rm rec}$ for a range of plasma magnetizations ($ 1 \le \sigma \le 30$) and obtained an approximate relation for $\sigma\gtrsim 10$ in electron-proton plasmas: $f_{\rm rec} \approx  0.25 \sigma/(\sigma + 2)$. For $\sigma \sim 1$ the dissipation efficiency decreases slightly ($\sim 0.1$). Moreover, for $\sigma \gtrsim 10$ the energy partition between relativistic protons and electrons is about the same, which justifies the use of the same $f_{\rm rec}$ for both species. Because of the linear dependence of $L'_{\rm e(p)}$ on $f_{\rm rec}$ and the fact that the efficiency changes only by a factor of 2.5 for the range of $\sigma$ values we consider, we adopt $f_{\rm rec}=0.25$ as a representative value in our calculations. 

Substitution of Eqs.~(\ref{eq:mdot}) and (\ref{eq:sigma2}) into Eq.~(\ref{eq:Lpart}) with $\beta \approx 1$ yields
\begin{eqnarray}
    L'_{\rm e(p)} & \simeq & 2.4 \times 10^{41} \frac{\sigma}{1+\sigma}\left(\frac{f_{\rm rec}}{0.25}\right) \left(\frac{\eta_{\rm j}}{0.9}\right)  \left(\frac{\eta_{\rm d}}{0.1}\right)^{-1}\nonumber \\ 
    & \cdot & \left(\frac{\dot{m}}{1.6\times10^{-5}}\right)^{1/3}L_{\rm Edd,46}~{\rm erg}\ {\rm s}^{-1}.
\end{eqnarray}
Given that $\dot{m} \propto \Gamma^3$ and $\mu = \Gamma (1+\sigma)$, we find that $L'_{\rm e/p}\propto \mu \sigma /(1+\sigma)^2$. Therefore, the relativistic particle luminosity is higher in jets with lower magnetizations and the same $\mu$. This will have an impact on the radiative output of blazar jets, as we will show in Sec.~\ref{sec:results}.

\subsection{Relativistic particle distributions}
Relativistic magnetic reconnection ($\sigma \ge 1$) is an efficient process for accelerating particles into broad power-law distributions~\citep[e.g.][]{Sironi_2014, Guo_2014, Werner_2016}. Kinetic simulations of relativistic reconnection in 2D and 3D have shown that the slope $p$ of the power-law distribution, $p = - {\rm d}\log N'/{\rm d}\log \gamma'$, depends on $\sigma$ with $p<2$ for $\sigma \gtrsim 10$, and $p>2$ otherwise ~\citep[e.g.][]{Sironi_2014, Guo_2014, Werner_2016}. Moreover, proton and electron distributions have similar slopes for $\sigma \gg 1$~\citep[e.g.][]{2016ApJ...818L...9G}. While this has not been clearly demonstrated for $\sigma \sim 1$ \citep[e.g.][]{2018MNRAS.473.4840W, 2019ApJ...880...37P}, we assume that $p_{\rm e}=p_{\rm p}=p$ for all values of $\sigma$ we study. We select indicatively $p \in \{3, 2.5, 2.2, 1.5, 1.2 \}$ for $\sigma \in \{1, 3, 10, 30, 50 \}$, while noting that differences of $\sim 0.3$ in the derived slopes between simulations are found.

Motivated by these results we model the volumetric injection rate of relativistic particles as
\begin{equation} \label{eq:18}
Q^{\rm inj}_{\rm i}(\gamma')\ = Q_{0, \rm i} {\gamma}'^{-p} \; \; \; {\rm for}\; {\gamma}'_{\rm i, min}<{\gamma}'<{\gamma}'_{\rm i, max} 
\end{equation}
where ${\gamma}'_{\rm i, min/max}$ are the minimum and maximum particle Lorentz factors, and $Q_{0, \rm i}$ is a normalization factor. This can be derived from the particle injection luminosity (see Eq.~(\ref{eq:Lpart})), and reads
\begin{equation} \label{eq:Qo}
Q_{0, \rm i} = \frac{4 L_{\rm B}f_{\rm rec}}{3{\beta}{\Gamma}^2 V' m_{\rm i}c^2 \mathcal{I}} 
\end{equation}
where $i={\rm e, p}$, $V^\prime = 4 \pi R^{'3}_{\rm b}/3$, and $\mathcal{I}$ is given by
$$
\mathcal{I}=\left\{\begin{array}{l}
\frac{\gamma_{\rm i, \max }^{\prime 2-p}-\gamma_{\rm i, \min }^{\prime 2-p}}{2-p} \;\;\; , p \neq 2 \\ \\
\ln \left(\frac{\gamma'_{\rm i, \max }}{\gamma'_{\rm i, \min }}\right)\;\;\;  , p=2
\end{array}\right.
$$ 

In the reconnection region the average energy per particle can be approximately written as $f_{\rm rec}\sigma m_{\rm p}c^2$. Thus, the mean Lorentz factor of particles with rest mass $m_{\rm i}$  is  
\begin{equation} \label{eq:gmean}
\langle \gamma'_{\rm i} \rangle \sim f_{\rm rec}\sigma\frac{m_{\rm p}}{m_{\rm i}}.
\end{equation}
The mean particle Lorentz factor of a power-law distribution with a finite energy range and slope $p \neq 2$ is also written as

\begin{equation}
\langle \gamma'_{\rm i} \rangle   = \frac{1-p}{2-p} \frac{\gamma_{\rm max, i}^{' -p+2}-\gamma_{\rm min, i}^{' -p+2}}{\gamma_{\rm max, i}^{' -p+1}-\gamma_{\rm min, i}^{' -p+1}}.
\label{eq:gmean2}
\end{equation}

For $p>2$ (i.e. $\sigma \lesssim 10$) and assuming $\gamma'_{\rm max, i} \gg \gamma'_{\rm min, i}$,  we can determine the minimum Lorentz factor of the distribution using Eqs.~(\ref{eq:gmean}) and (\ref{eq:gmean2}),

\begin{equation} \label{eq:gmin}
\gamma_{\rm min,i}^{\prime} \approx \frac{2-p}{1-p}{f_{\rm rec}} \sigma \frac{m_{\rm p}}{m_{\rm i}}.
\end{equation}
If the above expression yields values lower than 1, we set $\gamma_{\rm min, i}^\prime=1.25$.  
\begin{table*}
    \caption{Model parameters with their description and values.}
    \centering
    \begin{threeparttable}
    \begin{tabular}{lcccc}
    \hline 
     Parameter     &  Symbol & Value(s) \\
    \hline
     Input \\
     \hline
     Total energy flux normalized to rest-mass energy flux & $\mu$ & \{50, 70, 90\} \\
     Magnetization\tnote{*}     &  $\sigma$  & \{1, 3, 10, 30, 50\}\\
     Power-law index of accretion rate - bulk Lorentz factor relation & $s$ & 3 \\
     Disc radiative efficiency & $\eta_{\rm d}$ & 0.1 \\
     Ratio of jet power to accretion power & $\eta_{\rm j}$ & 0.9 \\
     Dissipated energy fraction transferred to relativistic particles & $f_{\rm rec}$ & 0.25 \\
     Power-law slope of particle distributions\tnote{**}  & $p$ & \{3, 2.5, 2.2, 1.5, 1.2\} \\ 
     Acceleration efficiency & $\eta_{\rm acc}$ & $10^3$ \\ 
     Minimum electron Lorentz factor & $\gamma_{\rm e,\min}$ & $10^3$ (for $p < 2$) \\
     Minimum proton Lorentz factor & $\gamma_{\rm p,\min}$ & $10^{0.1}$ (for $p<2$) \\
     Black hole mass & $M_{\rm BH}$ & $10^9 M_{\odot}$ \\
     BLR photon energy & $\epsilon_{\rm BLR}$ & 2 eV \\
     \hline 
     Derived \\
     \hline 
     Bulk Lorentz factor\tnote{$\dag$}    &  $\Gamma$ &  Eq.~(\ref{eq:1})\\ 
     Accretion rate normalized to the Eddington rate & $\dot{m}$ & Eq.~(\ref{eq:mdot})\\
     Total jet power & $L_{\rm j}$~(erg s$^{-1}$) & Eq.~(\ref{eq:Lj}) \\
     Injection luminosity of particle species $i$ & $L'_{\rm i}$~(erg s$^{-1}$) & Eq.~(\ref{eq:Lpart}) \\
     Minimum Lorentz factor of particle species $i$ & $\gamma_{\rm i,\min}$ & Eq.~(\ref{eq:gmin}) (for $p\ge 2$) \\
     Maximum Lorentz factor of particle species $i$ & $\gamma_{\rm i,\max}$ & Eq.~(\ref{eq:balance}) \\
     Blob radius & $R'_{\rm b}$~(cm) & Eq.~(\ref{eq:Rb}) \\
     Magnetic field strength of unreconnected plasma\tnote{$\ddag$} & $B'$~(G) & Eq.~(\ref{eq:B})\\
     Doppler factor & $\delta$ &  Eq.~(\ref{eq:delta}) \\
    \hline     
    \end{tabular}
    \begin{tablenotes}
\item[*] For $\mu=50$ we use $\sigma=48.9$. The magnetization $\sigma$ refers to the unreconnected plasma at the jet location where dissipation takes place. The magnetization of the reconnected plasma, however, is close to unity~\cite[e.g.][]{Sironi_2015, Hakobyan_2021}.
\item[**] In the same order as the values of $\sigma$ listed above.
\item[$\dag$] Computed for each pair of $(\mu, \sigma)$ values
\item[$\ddag$] It is taken to be the same as the magnetic field in the emitting region.
\end{tablenotes}
\end{threeparttable}
    \label{tab:param}
\end{table*}
The maximum Lorentz factor can be estimated through the balance of the acceleration and energy loss timescales of particles. Particles can accelerate via the reconnecting electric field, $E_{\rm rec} \sim \beta_{\rm rec} B'$, where $\beta_{\rm rec}\sim 0.1$ is the reconnection rate and $B'$ is the magnetic field strength of the unreconnected plasma\footnote{This is $\sim \sqrt{2}$ times lower than the average magnetic field in magnetic islands formed in the reconnection region \citep{Sironi_2016}.}. The characteristic acceleration timescale can be written as
\begin{equation} \label{eq:tacc}
t'_{\rm acc} \approx \frac{m_{\rm i} \gamma'_{\rm i}c^2}{e \beta_{\rm rec} B' c} = \eta_{\rm acc}\frac{r_{\rm g}}{c},
\end{equation}
where $r_{\rm g}$ is the gyroradius of a relativistic particle. Such fast acceleration (with $\eta_{\rm acc} \sim 10$) has been seen at the X-points of a current sheet in 2D simulations of reconnection \citep[see][and references therein]{Sironi_2022}.  Recently, \cite{Zhang_2021} demonstrated using 3D simulations of reconnection that non-trapped particles also undergo fast acceleration with $\eta_{\rm acc}\sim 10$. However, much slower acceleration processes of particles trapped within magnetic islands were also identified in large-scale 2D simulations \citep{Petropoulou_2018, Hakobyan_2021}. Being conservative we use $\eta_{\rm acc}=10^3$ in our numerical calculations\footnote{For comparison, \cite{2021MNRAS.501.4092R} adopted $\eta_{\rm acc}=10^{6}$, which resulted in smaller values of $\gamma'_{\max, \rm e}$ than those derived here.}. We then determine $\gamma'_{\rm max, i}$ by solving numerically the following equation,
\begin{equation}
    t'_{\rm acc} = \left(\sum_{j} t^{'-1}_{\rm loss, j}\right)^{-1}
    \label{eq:balance}
\end{equation}
taking into account all the relevant energy loss processes for each particle species.  While for electrons synchrotron radiation and inverse Compton scattering are the two competing energy loss processes in general, synchrotron cooling always dominates at the highest energies.  Protons could also lose energy via photopair and photopion production processes on jet photons and external radiation.  

Finally, for $1<p<2$ and $\gamma^\prime_{\rm max,i} \gg \gamma^\prime_{\rm min, i}$, we cannot determine anymore the minimum Lorentz factor from Eq.~(\ref{eq:gmin}). Instead, we use $\gamma_{\rm min,p}^\prime = 1.25$  and $\gamma^\prime_{\rm min, e}=10^3$ as indicative values. Our choice suggests that the power-law forms roughly above the proton rest-mass energy.

We summarize the model parameters in \tab{param} where we distinguish them in those used as an input to the numerical calculations and in those that are useful derived quantities.

\subsection{Numerical approach}\label{sec:code}
A useful approach for the study of non-thermal emission from time-variable astrophysical sources is the one involving solution of a system of differential equations describing the evolution of the radiating particle distributions (kinetic equation approach). 

The kinetic equations for a homogeneous emitting region containing relativistic particles of species $i$ can be cast in the following compact form

\begin{eqnarray}
\frac{\partial n'_{\rm i}(\gamma'_{\rm i}, t')}{\partial t'} & + &  \frac{n'_{\rm i}(\gamma'_{\rm i}, t')}{t'_{\rm esc, i}} + \sum_j\mathcal{L}^{\rm j}_{\rm i}(n'_{\rm i}, n'_{\rm k}, t')   =    \nonumber \\ 
& = & \sum_{\rm j}Q^{\rm j}_{\rm i}(n'_{\rm i}, n'_{\rm k}, t') + Q^{\rm inj}_{\rm i}(\gamma'_{\rm i},t'),
\label{eq:kinetic}
\end{eqnarray}
where $n'_{\rm i}$ is the differential particle number density, 
$t'_{\rm esc, i}=R'_{\rm b}/c$ is the particle escape timescale, $\mathcal{L}^{\rm j}_{\rm i}$ is the operator for particle losses (sink term) due to process $j$, $Q^{\rm j}_{\rm i}$ is the operator of particle injection (source term) due to process $j$, $Q^{\rm inj}_{\rm i}$ is the operator for the injection of accelerated particles, and index $i$ refers to protons (p), electrons/positrons (e), photons ($\gamma$), neutrons (n), and neutrinos ($\nu$). Note that the operators $Q^{\rm j}_{\rm i}$ and $\mathcal{L}^{\rm j}_{\rm i}$ of two-particle interactions (e.g. inverse Compton scattering, photopair and photopion production processes) generally depend on the densities of two particle species $i, k$. The coupling of the equations happens through the energy loss and injection terms for each particle species, and guarantees that the total energy lost by one particle species (e.g., protons) equals the energy transferred to other particles (e.g., pairs, neutrinos, and photons). 

\begin{figure}
    \centering
    \includegraphics[width=0.47\textwidth]{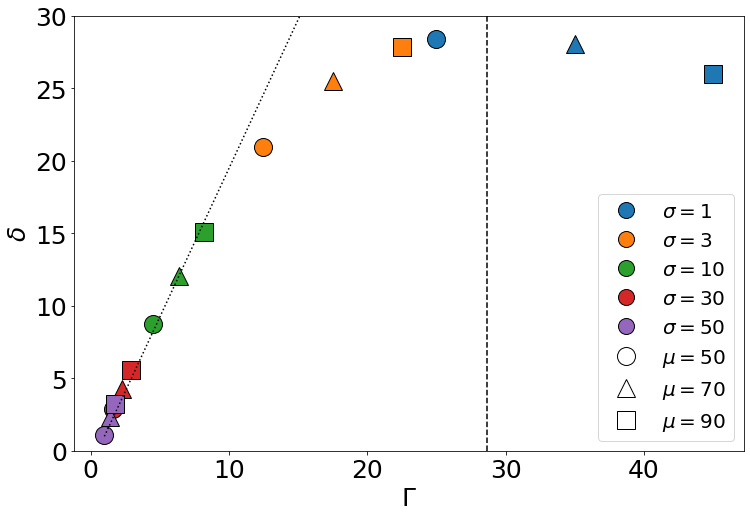}
    \caption{Doppler factor plotted against the jet bulk Lorentz factor for different values of $\mu$ and $\sigma$ as indicated in the legend, assuming $\theta_{\rm obs}=\pi/90$ (2~deg). The vertical dashed line marks the value $1/\theta_{\rm obs}$. The approximate relation $\delta=2\Gamma$ for $\Gamma \theta_{\rm obs}\ll 1$ is also shown (dotted line).}
    \label{fig:doppler}
\end{figure}

The main physical processes  that are included in Eq.~(\ref{eq:kinetic}) (for each stable species) are summarized below:
\begin{itemize}
    \item \textit{Electrons/positrons}: synchrotron radiation, inverse Compton scattering on synchrotron and external photons (using the full cross section), escape.
    \item \textit{Protons}: synchrotron radiation, photopair (Bethe-Heitler) production process, photopion production process, escape.
    \item \textit{Photons}: synchrotron radiation, synchrotron self-absorption, inverse Compton scattering, photon-photon pair production, neutral pion decay, escape.
    \item \textit{Neutrons}: photopion production process, escape.
    \item \textit{Neutrinos}: photopion production process, escape.
\end{itemize}
We compute the (comoving) photon and all-flavour neutrino energy spectra by solving the kinetic equations described in Eq.~(\ref{eq:kinetic}) until an equilibrium is reached (steady state), since we are not interested in the study of transient phenomena such as blazar flares. For the computations we use the numerical code {\sc athe$\nu$a} \citep{1995A&A...295..613M, 2012A&A...546A.120D}. Finally, we perform the appropriate transformations to obtain the spectra in the observer's frame using the Doppler factor,
\begin{equation}
    \delta = \frac{1}{\Gamma(1-\beta \cos(\theta_{\rm obs}))}
    \label{eq:delta}
\end{equation}
where we set $\theta_{\rm obs}=\pi/90$ (2~deg). The Doppler factor as a function of $\Gamma$ is plotted for reference in Fig.~\ref{fig:doppler} for 15 pairs of $(\mu, \sigma)$ values. Emission from jets with $\sigma \le 3$ will undergo the strongest Doppler beaming, which will impact the observed luminosity as we will show in the next section. Note that for $\Gamma \theta_{\rm obs} \gg 1$ or equivalently $\mu \theta_{\rm obs} \gg (1+\sigma)$ the Doppler factor begins to decrease. 

For completeness, we also list the parameter values used as input to the numerical code in tabular form in Appendix~\ref{sec:appA}.

\section{Results}\label{sec:results}

In this section we present the results for 15 simulated blazars that were obtained using the code described in Sec.~\ref{sec:code} for the parameter values listed in Table~\ref{tab:param}.  

\begin{figure}
    \centering
    \includegraphics[width=0.47\textwidth]{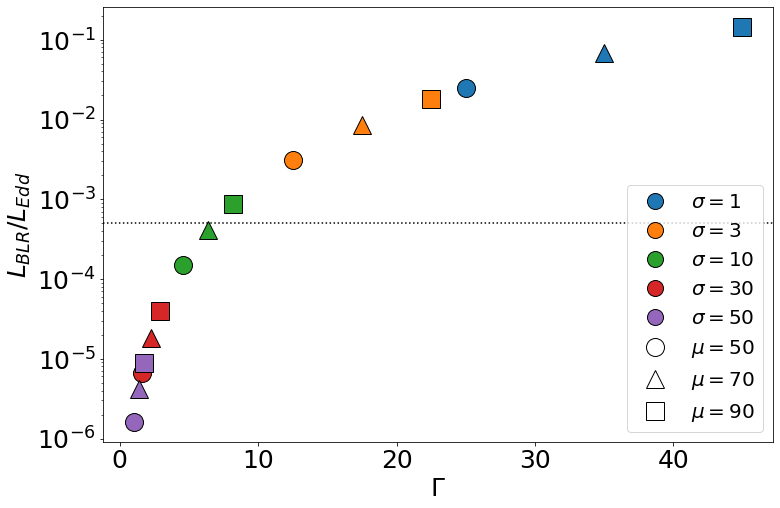}
    \caption{Ratio of the BLR luminosity and the Eddington luminosity (assuming $M=10^9 M_{\odot}$ and a BLR covering factor of 0.1) as a function of the jet Lorentz factor for all parameter sets we considered. The horizontal dotted line indicates the characteristic value that divides FSRQs from BL Lacs according to 
    \protect\cite{Ghisellini_2011}.}
    \label{fig:Lblr-Ledd}
\end{figure}

We begin by showing first the BLR luminosity (normalized to the Eddington luminosity) as a function of the jet Lorentz factor for all simulated blazars (see Fig.~\ref{fig:Lblr-Ledd}). For a fixed BLR covering fraction, as assumed here, and because of the adopted  $\dot{m}-\Gamma$ relation, we find that faster jets are associated with more luminous accretion discs and BLR emission. For comparison, we also indicate the value $L_{\rm BLR}/L_{\rm Edd}\sim 5\times10^{-4}$ that roughly divides FSRQs from BL Lac objects according to \cite{Ghisellini_2011}. This so-called \textit{blazar's divide} implies that differences between FSRQs and BL Lacs reflect differences in the accretion regime \citep[e.g.][]{Ghisellini_2009, Sbarrato_2014}. In what follows we will refer to simulated sources with $L_{\rm BLR}/L_{\rm Edd} \gg 5 \times10^{-4}$ ($\sigma \le 3$) as FSRQs, and as BL Lac objects otherwise ($\sigma > 10$). Therefore, a weak BLR is naturally present also in BL Lacs and can be used as a target photon field for both photohadronic interactions and inverse Compton scattering. Note that our results for $\sigma=10$ (green points in figure) fall in the transition regime.
We will also show in Sec.~\ref{sec:spectra} that the derived SEDs for $\sigma=10$ fall in between the high-luminosity and low-luminosity simulated blazars.

\begin{figure*}
    \centering
    \includegraphics[width=0.47\textwidth]{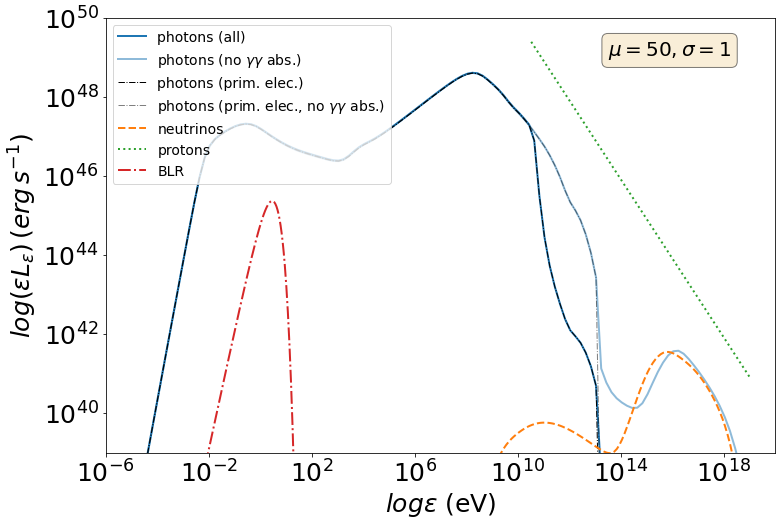}
    \hfill
    \includegraphics[width=0.47\textwidth]{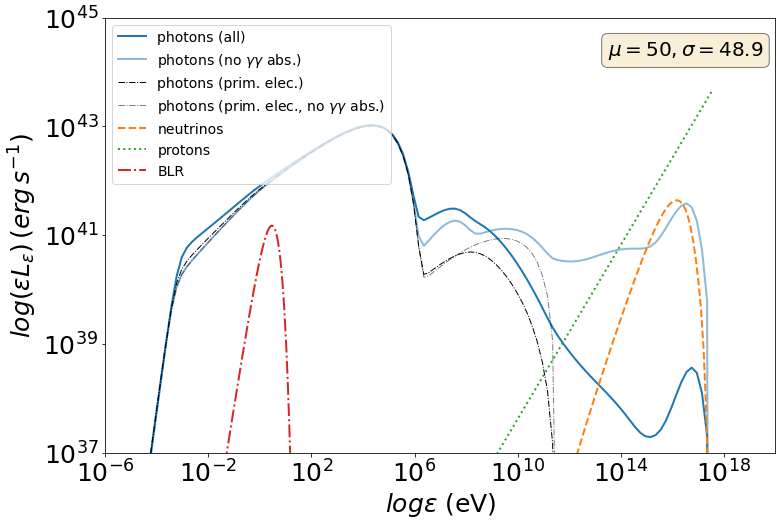}
    \caption{Broadband non-thermal photon spectrum (solid lines) and all-flavour neutrino spectrum (dashed lines) for $(\mu,\sigma)=(50,1)$ (left) and $(50,48.9)$ (right). The differential proton luminosity at injection (dotted line) and the BLR photon distribution in the AGN rest frame (dash-dotted line) are also shown. For comparison, the photon spectra produced by primary electrons are overplotted (dash-dotted line). Solid faint lines show the photon spectra without internal $\gamma \gamma$ absorption. No EBL $\gamma$-ray attenuation is included.}

    \label{fig:seds-1}
\end{figure*}
\subsection{Photon and neutrino spectra}\label{sec:spectra}
We continue by showing results of photon SEDs and neutrino spectra. To highlight the main differences in the spectra of simulated blazars with low and high magnetizations, we show first in Fig.~\ref{fig:seds-1} the results for  $\sigma=1$ (left) and $\sigma \sim 49$ (right) for $\mu=50$. In addition to the photon and neutrino spectra we also plot the proton  differential luminosity at injection, $\varepsilon_{\rm p} L_{\rm p}(\varepsilon_{\rm p}) = \delta^4 \varepsilon'_{\rm p} L'_{\rm p}(\varepsilon'_{\rm p})$ (where $\varepsilon_{\rm p} = \delta \varepsilon'_{\rm p}$ and $\varepsilon'_{\rm p}=m_{\rm p} \gamma'_{\rm p}c^2$), and the BLR photon field (in the AGN rest frame), which is approximated by a grey body of temperature $T_{\rm BLR} = \epsilon_{\rm BLR}/2.7 k_{\rm B}$ with energy density given by Eq.~(\ref{eq:uBLR-1}).

The SEDs of the two blazars differ not only in bolometric luminosity but also in shape. The blazar with $\sigma=1$ is several orders of magnitude more luminous than its high-$\sigma$ counterpart. Most of its energy radiated as GeV $\gamma$-rays, and its low-energy component peaks in the far infrared, which is consistent with the LSP classification \citep{PG_1995, Abdo_2010}. 
The disc emission, which is not shown here, would be brighter than the BLR emission by a factor of $\eta_{\rm BLR}^{-1}=10$. Still, it would remain hidden below 
the non-thermal emission of the low-$\sigma$ case. The high-energy component peaks at $\sim 1$~GeV and is composed of synchrotron-self Compton (SSC) and external Compton (EC) emission, with the former emerging as a small bump in the X-rays. These spectral characteristics are consistent with those of FSRQs \citep[see e.g.][]{Abdo_2010, Dermer_2014}. The high-$\sigma$ blazar, on the other hand, is less luminous, its broadband emission is synchrotron dominated, and has a $\sim 10$ keV peak synchrotron energy. These features are reminiscent of low-luminosity HSP BL Lac objects. The contribution of EC emission to the high-energy component is negligible because of the weak BLR emission  (not explicitly shown in the figure). 

The proton energy spectra (at injection) differ in these two cases by construction, since the power-law slope of particles accelerated via magnetic reconnection depends on $\sigma$ (see Table~\ref{tab:param}). As a result, most of the energy is carried by low-energy protons in the low-$\sigma$ case, as opposed to the high-$\sigma$ case, where most of the energy is carried by the most energetic particles of the distribution. Moreover, the total injection luminosity in protons (and electrons for that matter) is higher in the low-$\sigma$ blazar, since $L_{\rm e(p)} = \delta^4 L'_{\rm e(p)} \propto \mu^5 \sigma/(1+\sigma)^6$ assuming $\delta \sim \Gamma$ (see also Eq.~(\ref{eq:Lpart})).

In the low-$\sigma$ case, any hadronic-related emission (i.e. proton synchrotron radiation, secondary leptonic synchrotron and inverse Compton scattered radiation, and $\gamma$-rays from neutral pion decays), is not visible in the broadband photon spectra. In fact, the full spectrum (solid blue line) coincides with the one computed using emission from accelerated (primary) electrons (dash-dotted black line). Only when photon-photon absorption is omitted (for illustration purposes), does the hadronic contribution to very high-energy $\gamma$-rays ($\gtrsim 100$~TeV) become visible (compare solid and dash-dotted faint lines). These very energetic photons, which are produced from neutral pion decays, are attenuated in-source by lower energy photons. Moreover, 
the emission from the secondary pairs produced in this process is much less luminous than the primary leptonic emission, thus not altering the standard synchro-Compton spectrum. This is not true for the high-$\sigma$ case though. Looking first at the unattenuated photon spectrum (solid faint blue line), we see two bumps in the range of 1 MeV to 100 GeV. The MeV peak is attributed to proton synchrotron radiation, which becomes visible due to the combination of the strong magnetic field in the emitting region, a  high value of $\gamma^\prime_{\rm max, p}$  (see Table~\ref{tab:mu50}) and the hard proton spectrum (dotted green line) -- see also Model B in \cite{2017MNRAS.464.2213P} for similar results. The second bump peaking at $\sim 10$~GeV is produced by the SSC emission of primary electrons (see dash-dotted faint black line). Finally the bump at $\sim 10$~PeV is the result of neutral pion decays. In this case, the hadronic-related spectral components have comparable luminosity to the primary Compton emission. As a result, the emission from secondaries produced by $\gamma \gamma$ pair production modifies the $\gamma$-ray spectrum at $\gtrsim 1$~MeV washing out the two bumps, and changes slightly the primary leptonic synchrotron component at energies $\lesssim 1$~eV. 

While the electromagnetic signatures of the hadronic component are in most cases not visible, high-energy neutrinos are free streaming from the source upon their production without undergoing any attenuation. In both cases, the neutrino emission peaks at $\sim 10$~PeV. However, there are two important differences between the low-$\sigma$ and high-$\sigma$ cases. First, the neutrino-to-$\gamma$-ray luminosity ratio 
is much smaller in the  low-$\sigma$ case than in the high-$\sigma$ case, suggesting a progressively more important role of the photopion process in the source as $\sigma$ increases (see also Sec.~\ref{sec:Y-xi}). Second, the neutrino spectrum in the low-$\sigma$ blazar shows two bumps, a more luminous one peaking at $\sim 10$~PeV, and a less luminous one peaking at $\sim 100$~GeV. The lower energy peak of the neutrino spectrum is related to lower energy protons that interact with high-energy non-thermal photons, while the higher energy bump is attributed to interactions with the BLR photons. As $\sigma$ increases the number density of BLR photons decreases, namely $n'_{\rm BLR}\approx u'_{\rm BLR}/\epsilon'_{\rm BLR} \propto \mu/(1+\sigma)$ (see also Eq.~\ref{eq:uBLR}). As a result, non-thermal jet photons become the main target for protons in the high-$\sigma$ case. We refer the interested reader to Appendix~\ref{sec:appB} for a semi-analytical derivation of the neutrino spectra that qualitatively explains this trend.

\begin{figure*}
    \centering
    \includegraphics[width=0.47\textwidth]{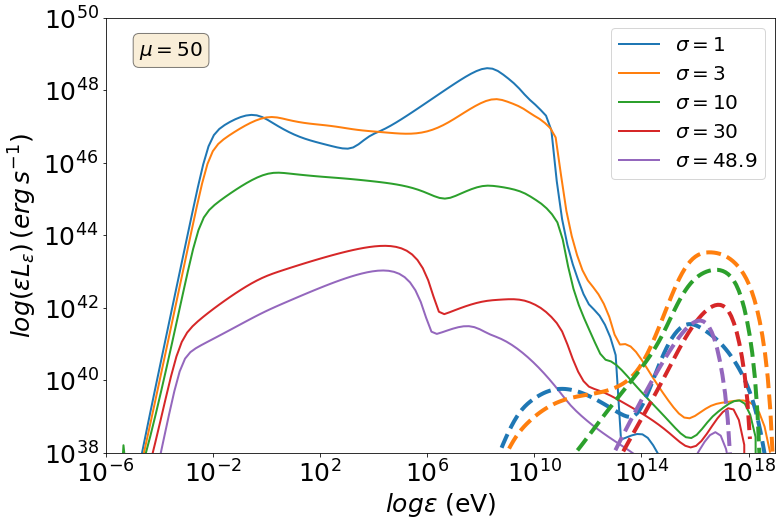}
    \hfill
    \includegraphics[width=0.47\textwidth]{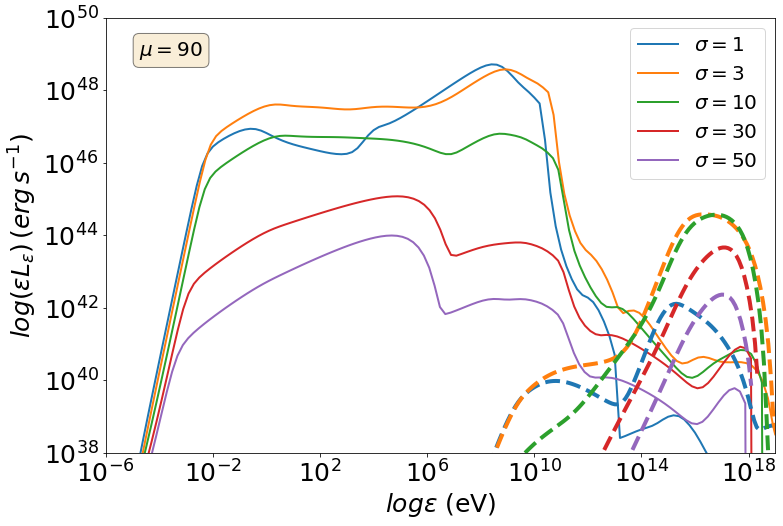}
    \caption{Broadband non-thermal photon spectra (thin solid lines) and all-flavour neutrino spectra (thick dashed lines) for $\mu=50$ (left) and 90 (right) for different magnetizations (see inset legends). No EBL $\gamma$-ray attenuation is included.}
    \label{fig:seds-2}
\end{figure*}

Figure~\ref{fig:seds-2} shows the photon and neutrino spectra obtained for $\mu=50$ (left) and 90 (right) for various magnetizations. Similar trends are found for $\mu=70$ and therefore the respective plot is omitted. We comment first on the $\mu=50$ results and then make a comparative discussion of the results for the two $\mu$ values. 

The photon spectra of the simulated blazars become more luminous for lower magnetizations in agreement with \cite{2021MNRAS.501.4092R}. This trend can be understood by the combined result of two factors. First, the Doppler boosting is stronger for lower $\sigma$ values, as illustrated in Fig.~\ref{fig:doppler}. Second, the particle injection luminosity (in the comoving frame) also decreases with $\sigma \gtrsim 3$, since $L'_{\rm e(p)} \propto \mu \sigma/(1+\sigma)^2$ (see Eq.~(\ref{eq:Lpart})). In all cases, the low-energy hump is attributed to primary electron synchrotron radiation. The high-energy hump is either explained by EC scattering off BLR photons by the jet electrons for $\sigma \leq 10$, or it is attributed to SSC for $\sigma=30$, or it is a combination of secondary leptonic emission and  SSC of primary electrons for $\sigma=50$. 

Overall, we do not find strong evolution of the peak synchrotron energy with $\gamma$-ray luminosity for the simulated blazars that belong either in the high-luminosity group ($\sigma < 10$) or the low-luminosity group ($\sigma \ge 30$). Moreover, there is a clear increase of the Compton dominance with $\gamma$-ray luminosity in the high-luminosity group, which is broadly consistent with the \fermi \, (revised)  blazar sequence~\citep{Ghisellini_2017,2022Galax..10...35P}. The simulated blazar for $\sigma=10$ is an outlier from the known spectral subclasses, as it has an almost flat synchrotron spectrum that spans about 6 orders of magnitude in energy. Its peculiar spectrum is the combined result of (i) rough energy equipartition between magnetic fields and BLR photons, and (ii) fast cooling electrons injected with $p=2.2$ and a high maximum Lorentz factor ($\sim 10^6$). To better illustrate the differences in the radiating particle distributions, we present the steady-state electron energy distributions for the displayed photon spectra in Appendix~\ref{sec:appC}. 

A major difference between the non-thermal photon and neutrino emission is that the neutrino luminosity is not a monotonic function of the magnetization. Instead, we obtain the most luminous neutrino emission for $\sigma=3$. This more complex behaviour arises from an additional factor that needs to be taken into account when computing the neutrino spectra. Besides the target photon density (jet or BLR photons), the Doppler factor, and the bolometric proton injection luminosity, one has to consider the energy threshold of the interactions, and the integrated proton luminosity from this threshold and above (see Appendix~\ref{sec:appB} for more details).

Jets launched with higher $\mu$ values are more luminous, but with similar spectral shapes. This trend is partially explained by the fact that the injection electron luminosity (in the observer's frame) scales as $L_{\rm e(p)} \propto \mu^5\sigma/(1+\sigma)^6$. If the electrons are radiating away their energy efficiently, the photon luminosity should follow a similar scaling with the injection luminosity. 
Notice also the change in the Compton dominance (i.e. the luminosity ratio of the Compton and synchrotron components) between the low and high $\mu$ values for $\sigma=1$, which implies a higher ratio $u'_{\rm BLR}/u'_{\rm B}$ for $\mu=90$. Indeed, combination of Eqs.~(\ref{eq:uBLR}) and (\ref{eq:B})  yields $u'_{\rm BLR}/u'_{\rm B} \propto \mu^2/[\sigma (1+\sigma)]$. This scaling also highlights the progressively diminishing role of EC scattering in the SEDs of more strongly magnetized jets, in agreement with the findings of \cite{2021MNRAS.501.4092R}. With the exception of the Compton dominance, $\mu$ does not have a strong impact on the spectral shape. This is mostly due to the fact that the properties of the radiating particles (e.g. $p$ and $\gamma'_{\rm min, e}$) and the magnetic field strength are independent of $\mu$. Finally, the neutrino spectra for $\mu=90$ are more luminous than those for $\mu=50$, while having similar shapes for all values of $\sigma$ we considered.

\subsection{Baryonic loading and neutrino-to-$\gamma$-ray luminosity ratio}\label{sec:Y-xi}
\begin{figure}
    \centering
    \includegraphics[width=0.47\textwidth]{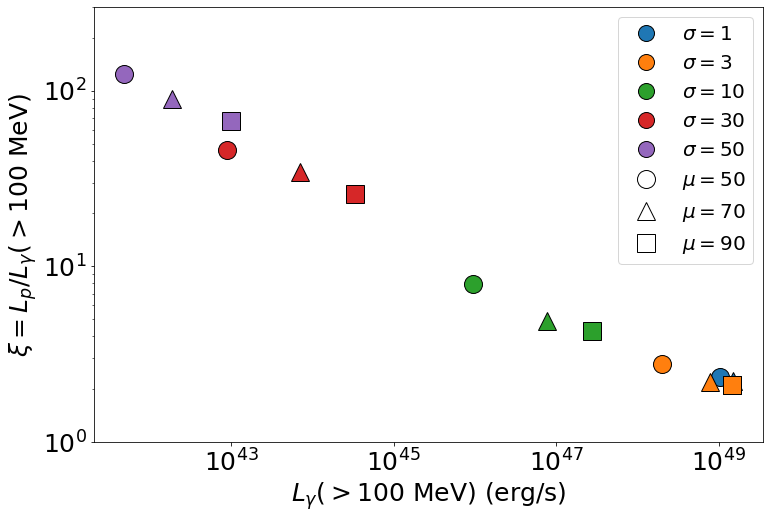}
    \caption{Baryonic loading of the simulated blazars plotted against the observed $\gamma$-ray luminosity integrated above 100~MeV. Different markers are used for different $(\mu, \sigma)$ values (see inset legend).}
    \label{fig:xi-Lg}
\end{figure}

The expected (all-flavour) neutrino luminosity of a blazar can be estimated as 
\begin{equation}
    L_{\nu+\bar{\nu}} \approx  \frac{3}{8}f_{\rm p\pi} L_{\rm p} = \frac{3}{8} f_{\rm p\pi} \xi L_{\gamma} = Y_{\nu \gamma} L_{\gamma}
    \label{eq:Lv-Lg}
\end{equation}
where $f_{\rm p\pi}=1/(1+t'_{\rm p\pi}/t'_{\rm dyn})$ is the so-called photopion production efficiency, $t'_{\rm p\pi}$ is the respective energy loss timescale for protons, $t'_{\rm dyn}\sim R'_{\rm b}/c$ is the source dynamical timescale, $L_{\rm p}=\delta^4 L'_{\rm p}$, and $\xi \equiv L_{\rm p}/L_{\gamma}$ is known as the baryonic loading of the source. 
The latter is largely unknown and it is taken as a free parameter in most leptohadronic models~\citep[e.g.][]{Murase_2014, Petropoulou_2015}, with the exception of models where $\gamma$-rays are explained by proton synchrotron radiation~\citep[e.g.][]{Cerruti_2015, PD_2015, PD_2016}. The photopion efficiency also depends on various physical parameters, such as the Doppler factor and size of the source~\citep[for explicit expressions, see e.g.][]{Murase_2014, PM_2015}. Fig.~\ref{fig:xi-Lg} shows the baryonic loading for the simulated blazars as a function of the $\gamma$-ray luminosity\footnote{Here, $L_{\gamma}$ represents the integrated $\gamma$-ray luminosity at energies above 100~MeV.}. The obtained $\xi$ values lie in the range of 1.5 and 150, showing a very weak decrease with increasing $L_{\gamma}$. The baryonic loading is not as extreme as those obtained from SED modelling of individual blazars that aim to maximize their neutrino output~\citep[see e.g.][]{Petropoulou_2020}. 
It is interesting to note that in our model where $L_{\rm p}=L_{\rm e}$ the baryonic loading is also equivalent to the inverse of the $\gamma$-ray efficiency  $L_{\gamma}/L_{\rm e}$  (at least for the low-$\sigma$ jets where the hadronic contribution to the $\gamma$-ray emission is negligible).

\begin{figure}
    \centering
    \includegraphics[width=0.47\textwidth]{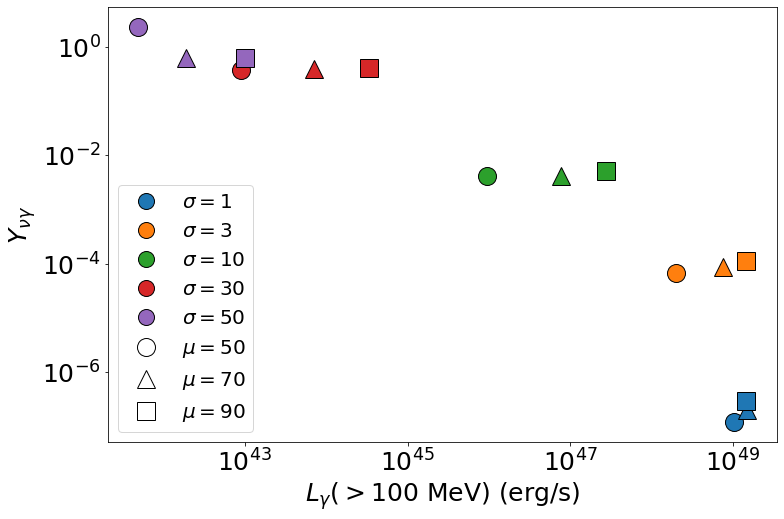}
    \caption{Ratio $Y_{\nu\gamma}$ plotted against the $\gamma$-ray luminosity integrated above 100~MeV for all simulated blazars. Different markers are used for different $(\mu, \sigma)$ values (see inset legend).}
    \label{fig:Y-Lg}
\end{figure}

All the information carried by the photopion efficiency and the baryonic loading can be incorporated into a single parameter, $Y_{\nu \gamma}$, that is the ratio of the all-flavour neutrino luminosity to the observed $\gamma$-ray luminosity of a blazar~\citep[e.g.][]{Petropoulou_2015, Palladino_2019}. In Fig.~\ref{fig:Y-Lg} we plot this ratio against the $\gamma$-ray luminosity  for 15 pairs of $(\mu, \sigma)$ values. We find a decreasing trend of $Y_{\nu \gamma}$ with increasing $L_{\gamma}$  for $\sigma \le 10$, which is mostly driven by the strong dependence of $L_{\gamma}$ on $\sigma$, and to a lesser degree by the variation of $L_{\nu}$ with $\sigma$ (see also Fig.~\ref{fig:seds-2}). Moreover, the ratio is not affected much by the value of $\mu$. We find $Y_{\nu \gamma} \ll 0.1$ for the FSRQ-like simulated sources and the transitional blazar (i.e. for $\sigma \le 10$), while $Y_{\nu \gamma} \sim 0.3-1$ for the BL Lac-like simulated blazars ($\sigma=30,50$).

Qualitatively similar findings were reported in earlier studies using very different approaches \citep{Palladino_2019, Petropoulou_2020}. 
More specifically, \cite{Petropoulou_2020} gathered results about $Y_{\nu \gamma}$ from leptohadronic models applied to various BL Lac objects with $L_{\gamma}\gtrsim 10^{45}$~erg s$^{-1}$ (see their Fig.~15), including the extreme HSP blazar 3HSP J095507.9+355101~\citep{2020A&A...640L...4G, Paliya_2020}, six candidate neutrino sources from earlier works~\citep{Petropoulou_2015}, and the masquerading BL Lac TXS~0506+056~\citep{Keivani2018, 2020ApJ...891..115P}. These ratios were obtained by modelling of the SEDs using the highest proton luminosity allowed by the observations in each case. In these models, however, physical quantities, such as the electron and proton luminosities, the power-law slopes, the BLR luminosity, the Doppler factor, and others were treated as free parameters, unlike in our model where all of them are ultimately related to $\mu$ and $\sigma$. \cite{Palladino_2019} explored different scenarios for the baryonic loading of blazars while trying to explain the diffuse neutrino flux in terms of the blazar sequence. They concluded that a scenario where $\xi$ and $Y_{\nu \gamma}$ are anti-correlated with $L_{\gamma}$ is plausible (see their Fig.~8). According to this, low-luminosity BL Lacs should be bright neutrino
sources to power the IceCube neutrino flux, while 
FSRQs should be dim in neutrinos. Our results for the individual simulated blazars are qualitatively similar with these previous findings, while providing a physically motivated framework.

\begin{figure}
    \centering
    \includegraphics[width=0.47\textwidth]{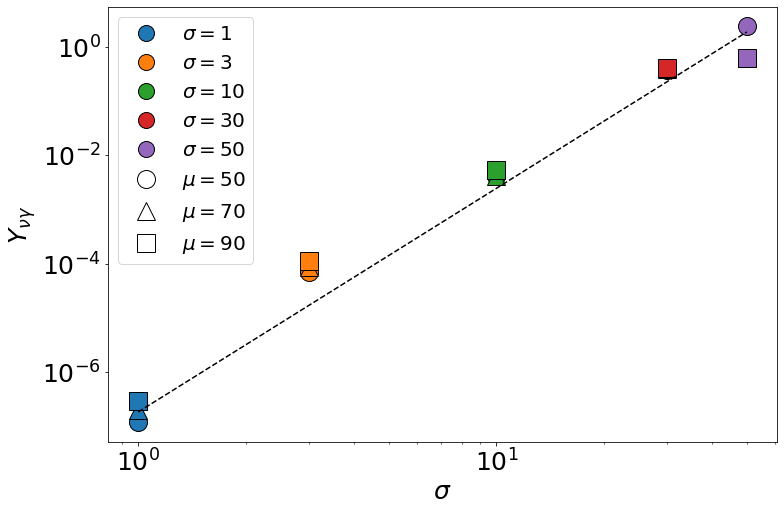}
    \caption{Ratio $Y_{\nu \gamma}$ as a function of $\sigma$ for all simulated blazars. The trend can be approximated by the expression in Eq.~(\ref{eq:Y-approx}) (dashed line).}
    \label{fig:Y-sigma}
\end{figure}

In Fig.~\ref{fig:Y-sigma} we also show the ratio $Y_{\nu \gamma}$ as a function of $\sigma$. Because of the strong dependence of $L_{\gamma}$ on $\sigma$, the ratio increases for higher magnetizations. This trend can be approximated by the following expression,
\begin{equation}
\log \left( Y_{\nu \gamma} \right) \approx -6.7 + 4.1 \cdot \log(\sigma) 
\label{eq:Y-approx}
\end{equation}
which is overplotted in Fig.~\ref{fig:Y-sigma} with a dashed line.
Different values of $\mu$ produce only a small scatter around the predicted value from the expression above. Combining the information about $Y_{\nu \gamma}$ and $\xi$ we can also infer the photopion production efficiency. For example, for high-$\sigma$ sources where $Y_{\nu \gamma}\sim 1$ and $\xi\sim 100$, we can estimate using Eq.~(\ref{eq:Lv-Lg}) that $f_{\rm p\pi}\sim 0.01$, which is also verified by semi-analytical calculations (see Appendix \ref{sec:appB}).

\subsection{Jet power}
\begin{figure}
    \centering
    \includegraphics[width=0.47\textwidth]{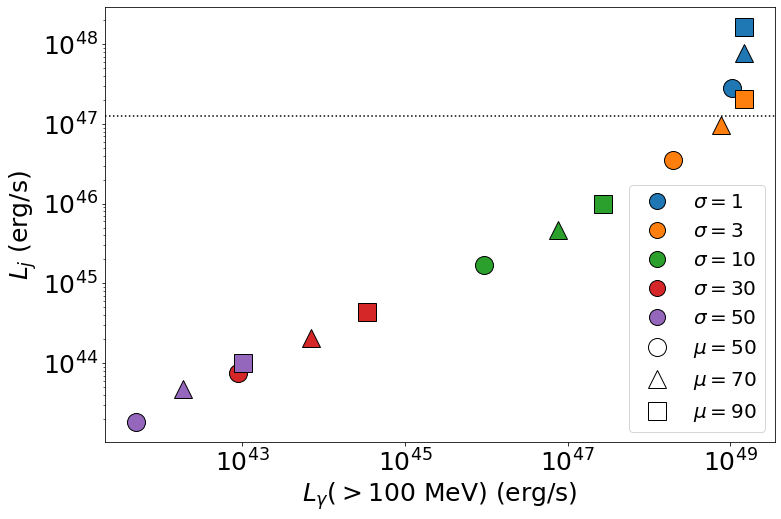}
    \caption{Jet power plotted against the observed $\gamma$-ray luminosity ($>100$~MeV) for all simulated blazars. Different markers are used for different $(\mu, \sigma)$ values (see inset legend). The Eddington luminosity for a $10^9 M_{\odot}$ black hole is also marked for reference (dotted line).}
    \label{fig:Lj-Lg}
\end{figure}

The jet power, $L_{\rm j}$, is plotted against the observed $\gamma$-ray luminosity (integrated above 100 MeV) for all simulated blazars in Fig.~\ref{fig:Lj-Lg}. Being proportional to ${\Gamma}^3$, $L_{\rm j}$ increases for lower magnetizations (see Eqs.~(\ref{eq:1}), (\ref{eq:mdot}) and (\ref{eq:Lj})).  For comparison reasons, a horizontal line which stands for the Eddington luminosity of a supermassive black hole with of $10^9M_{\odot}$ is also plotted in the same figure. All the low-luminosity blazars in our model (i.e. for $\sigma>10$) have $L_{\rm j}\ll L_{\rm Edd}$. This makes our model energetically favourable compared to other leptohadronic models for BL Lac objects presented in the literature that require super-Eddington jet power~\citep[e.g.][]{Petropoulou_2015, Petropoulou_2020}. Still, the high-luminosity blazars, which correspond to $\sigma \le 3$ and $\dot{m} \gtrsim 10^{-1}$ (see also Fig.~\ref{fig:Lblr-Ledd}), have jet power close to or even exceeding $L_{\rm Edd}$ in agreement with previous findings \citep[e.g.][]{Ghisellini_2014}. 

\section{Discussion}\label{sec:discussion}
In this paper we have expanded the work of \cite{2021MNRAS.501.4092R} by computing the photon and high-energy neutrino emissions produced by the interactions of relativistic protons with photons. While the electromagnetic signal of proton-related processes (e.g., proton synchrotron radiation, synchrotron radiation of pairs from charged pion decays and photopair production) is in most cases hidden below the emission from primary electrons, the associated neutrino signal peaks at a few PeV, making it relevant for current and future neutrino detectors, like IceCube, IceCube-Gen2~\citep[][]{IceCube-Gen2,Aartsen_2021gen2} and KM3Net \citep[][]{aiello2019arca}.

Our findings about the blazar SED are in general agreement with those presented in \cite{2021MNRAS.501.4092R}. There are a couple important differences, however, that are worth mentioning. First, electron cooling in the Klein-Nishina regime is included in our calculations (with an approximate way as described in \citealt{1995A&A...295..613M}). This becomes important in the low-$\sigma$ regime, when $\Gamma$ becomes large and the BLR photons (which are the main seed photons for inverse Compton scattering) are boosted to higher energies. Because of the less efficient cooling of electrons in the Klein-Nishina regime, the cooled part of the distribution does not have the standard slope of $-p-1$, but it is flatter (see Appendix~\ref{sec:appC}). Therefore, the synchrotron spectra for $\sigma=1$ and 3 are not as steep as those presented in \cite{2021MNRAS.501.4092R}. Moreover, the injected electron distributions for $\sigma < 10$ (where $p>2$) extend to higher Lorentz factors than in \cite{2021MNRAS.501.4092R}, because we adopted a lower value of $\eta_{\rm acc}$ which is closer to the one expected in reconnection. Furthermore, we find no strong evolution of the synchrotron peak energy with $L_{\gamma}$ for the low-luminosity simulated blazars ($\sigma \ge 30$) contrary to \cite{2021MNRAS.501.4092R} (see their Fig.~1). This stems from the different way of defining $\gamma^\prime_{\rm max, e}$, which will be discussed in more detail later in this section. Besides these differences in implementation, the blazar SEDs shown in Fig.~1 of \cite{2021MNRAS.501.4092R} are less luminous than those shown in our Fig.~\ref{fig:seds-2} for the same $(\mu, \sigma)$ values. This can be explained by the lower value of the reconnection efficiency used in \cite{2021MNRAS.501.4092R} and the time-averaging of the displayed spectra in their paper.

The new element of this work is the calculation of the expected neutrino emission from FSRQ-like and BL Lac-like blazars in a common framework. We showed that the all-flavour neutrino spectrum peaks at $\sim 10$~PeV for all values of $\sigma$, with a total luminosity being weakly dependent on $\sigma$ (see Fig.~\ref{fig:seds-2}).
We computed, $Y_{\nu \gamma}\equiv L_{\nu + \bar{\nu}}/L_{\gamma}$, for all values of $(\mu, \sigma)$ and showed that $Y_{\nu \gamma}\propto \sigma^4$. As the low-luminosity blazars in our model are associated with high magnetization, they have $Y_{\nu \gamma} \sim 0.3-1$. Ratios close to unity suggest a significant hadronic contribution to the $\gamma$-ray spectrum above 100 MeV (see e.g. Fig.~\ref{fig:seds-1}). The ratio $Y_{\nu \gamma}$ has been constrained by IceCube using different methods. A fit to the diffuse astrophysical neutrino flux with a composite spectral model that accounts for the contribution of HSP BL Lacs based on the model of  \cite{Padovani_2015} yielded $Y_{\nu \gamma} < 0.41$ at 90\% C.L. \citep{Aartsen_2020b}. The strictest upper limit on $Y_{\nu \gamma}$ so far is 0.13 and is placed by the IceCube non detections at ultra-high energies \citep{Aarsten_2016, Aartsen_2017b}. While the highest ratios obtained here appear  inconsistent with the strictest upper limit, one should note that the latter was derived  based on the model of \cite{Padovani_2015} where the peak neutrino energy was related to the synchrotron peak frequency and a common value of $Y_{\nu \gamma}$ for all HSPs was assumed. Still, a more careful comparison of our model to existing ultra-high energy upper limits of IceCube is warranted.  

Recently, \cite{Giommi_2020} presented a sample of 47 $\gamma$-ray selected ISP and HSP blazars, out of which about 16 could be associated with individual neutrino track events detected by IceCube. Follow-up spectroscopy of these sources and use of multi-frequency diagnostics revealed that masquerading BL Lacs consist more than $24$ percent of the sample~\citep{SIN2}. They typically have $ 10^{45} < L_{\gamma}(>100~\rm MeV) < 10^{47}$~erg s$^{-1}$ (see Figs.~2 and 3 in \citealt{SIN2}). Our simulated blazars for $\sigma=10$ fall in this range of $\gamma$-ray luminosities and are characterized by $Y_{\nu \gamma} \sim 4\times 10^{-3}$. Moreover, the simulated blazars for $\sigma=10$ fall in the transition region between FSRQs and BL Lacs in terms of their BLR luminosity (see Fig.~\ref{fig:Lblr-Ledd}). Hence, we could tentatively compare them to masquerading BL Lacs. It is noteworthy that TXS~0506+056~\citep{Padovani_2019} and PKS~0735+178 \citep{Sahakyan2022} are also masquerading BL Lacs that have been associated with high-energy neutrinos. Leptohadronic modelling of these sources resulted in values (or upper limits) of $Y_{\nu \gamma}$ in the range $0.003-0.03$~\citep[see e.g.][]{Petropoulou_2020, Sahakyan2022}. These findings are consistent with those of our work, if one considers a possible scatter around the $Y_{\nu \gamma}$ values presented in Fig.~\ref{fig:Y-Lg} (due to e.g. different black hole masses). Even higher values of $Y_{\nu \gamma}$ for these intermediate blazars can be obtained in our model, if the proton power-law slope for $\sigma=10$ is closer to 2. On the contrary, the high-luminosity (FSRQ-like) sources in our model are deemed to be dim in neutrinos, because the proton distribution is very soft ($p\sim 3$) for $\sigma \gtrsim 1$.

In our model, neutrinos are produced via interactions of relativistic protons with photons produced locally via leptonic processes, and externally provided by the BLR. Inelastic proton-proton (pp) collisions are another likely process for neutrino production. These could take place between the relativistic proton population and the cold proton plasma of the jet. To estimate the importance of pp collisions we can compare the respective efficiency with that of the photopion production process.  Combining the definition of  magnetization, from Eq.~(\ref{eq:sigma1}), with Eqs.~(\ref{eq:Rb}), (\ref{eq:Lj}), and (\ref{eq:B}), we may write the cold proton density as
\begin{eqnarray}
n_{\rm p, c}^{\prime} \simeq \frac{7.8 \times 10^3}{(1+\sigma)\beta} \left(\frac{\eta_{\rm j}}{0.9}\right) \left(\frac{\eta_{\rm d}}{0.1}\right)^{-1} \;\; \mathrm{cm^{-3}}.
\end{eqnarray}
Assuming a constant effective cross section for the pp process, $\hat{\sigma}_{\rm pp} = k_{\rm pp} \sigma_{\rm pp}\approx 25 \times 10^{-27}$~cm$^2$~\citep{DM_2009}, we can estimate the efficiency of the $pp$ process, as 
\begin{equation}
   f_{\rm pp} \approx n'_{\rm p, c} \hat{\sigma}_{\rm pp} R'_{\rm b} \simeq 2.5\times10^{-7}  n'_{\rm p, c, 4} R'_{\rm b, 15}.
   \label{eq:fpp}
\end{equation}
For comparison, the efficiency of the photopion production process is $f_{\rm p \pi} \gtrsim 10^{-3}$ for interactions with BLR photons  and $\gtrsim 10^{-6}-10^{-2}$ for high-energy proton interactions with jet photons (see Appendix~\ref{sec:appB}). Therefore, pp collisions are a negligible source of neutrinos for energies above the photopion production threshold. Nevertheless, they can contribute to the neutrino spectrum at lower energies (i.e. at tens of GeV to TeV energies), but at a much lower luminosity than the one depicted in Fig.~\ref{fig:seds-2}.

The predicted neutrino emission for the low-luminosity simulated blazars ($\sigma\ge 30$) depends strongly on the maximum proton Lorentz factor. In this work, we determined the latter by balancing the acceleration rate with the total energy loss rate. This approach for determining the maximum Lorentz factor assumes that a steady state is achieved for the highest energy protons. This, however, might not be always true. In this case, the maximum Lorentz factor would be determined by equating the dynamical (expansion) time of the emission region (blob) with the acceleration timescale. This approach would yield a lower $\gamma'_{\rm max, p}$ than the one used in this work. This can be understood if we consider that (i) $t_{\rm loss,p} \sim (10-100) \times R'_{\rm b}/c$ even for the high-energy protons in all cases we explored, and (ii) the  dynamical timescale is expected to be a few times longer than the light-crossing time of the emitting region $R'_{\rm b}/c$.

For the high-$\sigma$ cases where hard power-law proton distributions are expected ($1<p<2$), we could estimate the maximum Lorentz factor in a different manner (adopted also in \cite{2021MNRAS.501.4092R}). Using  Eqs.~(\ref{eq:gmean}) and (\ref{eq:gmean2}) and taking into account that the mean energy per particle cannot exceed by a lot $(\sigma +1) m_{\rm p}c^2$ due to energetic constraints ~\citep[e.g.][]{Werner_2016}, one can express the maximum particle Lorentz factor as 
\begin{equation}\label{eq:gmax}
\gamma_{\rm max,i}^{\prime}=\left[\frac{2-p}{p-1} \operatorname{f_{\rm rec}}(\sigma+1) \frac{m_{\rm p}}{m_{\rm i}}\right]^{1 / (2-p)} \quad \!\!\! \gamma_{\rm min,i}^{\prime (1-p) / (2-p)}.
\end{equation}
When applied to electrons, the expression above predicts for $p<2$ an evolution of the peak synchrotron energy with $\sigma$, similar to the one presented in \cite{2021MNRAS.501.4092R}. Direct application of the above equation to protons yields small maximum Lorentz factors (e.g. $\gamma^\prime_{\rm max, p}\sim 162$ for $\sigma=50$ and $\sim 6.3\times 10^4$ for $\sigma=1000$ and $p=1.5$). In this case the energy threshold for photomeson interactions on the jet synchrotron photons or the BLR photons is not satisfied, making inelastic pp collisions  (see previous paragraph) and photopion production on low-density Compton-scattered photons the only relevant mechanisms for neutrino production. As a result, low-luminosity HSP BL Lacs would also be dim neutrino sources, thus making the perspectives of detecting individual blazars in neutrinos poor. We note, however, that if there is a progressive softening of the power-law slope with time to $p\sim 2$ even for high $\sigma$ values, then Eq.~(\ref{eq:gmax}) is not a hard limit. Some hints for this process have been reported by \cite{Petropoulou_2018, Hakobyan_2021}, but the evolution was seen on long timescales (i.e. several light crossing times of the reconnection layer). An asymptotically softer proton spectrum for the high-$\sigma$ cases ($p\sim 2$) would result in lower peak neutrino energy and luminosity than those presented here, since most of the energy stored in relativistic protons would be carried by the lower energy protons of the distribution.  A more careful analysis of these effects is however beyond the scope of this work.

An important assumption of our work has to do with the location of the emitting region, which was fixed at the edge of the BLR. As a result, in all simulated blazars, the BLR photon field appeared boosted in the comoving frame of the emitting blob in the jet~\citep[e.g.][]{Ghisellini_Madau_1996}. This assumption is crucial for the low-$\sigma$ blazars, where BLR photons are the main seed for inverse Compton scattering. If the emitting region was located well beyond the BLR \citep[see e.g.][]{Costamante_2018}, then the BLR number density would appear de-boosted~\citep[e.g.][]{Dermer_1994}, and the EC component for $\sigma \le 3$ would be suppressed. Similarly, the neutrino production rate would decrease, thus reducing the high-energy peak of the neutrino spectrum. As a result, the low-$\sigma$ sources would become even dimmer in neutrinos. At distances beyond the BLR, but still within the dusty torus, which is located at $\sim$pc scales \citep[e.g.][]{2000ApJ...545..107B}, infrared (IR) photons from the torus become the relevant targets for EC scattering and photopion production~\citep[for an application to PKS~1502+106, see][]{Oikonomou_2021, 2021ApJ...912...54R}. In this case, the peak neutrino energy would shift to higher values, because a higher proton energy would be needed to satisfy the pion-production threshold on IR photons compared to the BLR photons\footnote{The energy threshold condition for head-on proton-photon collision reads $2 \gamma'_{\rm p} \epsilon' \gtrsim 145$~MeV, where $\epsilon'$ is the target photon energy.}. As far as the peak neutrino luminosity is concerned, one has to estimate the pion production efficiency on IR photons and the proton luminosity above the energy threshold, as shown for the BLR photons in Appendix~\ref{sec:appB}. First, the ratio of photon number densities can be written as 
\begin{equation}
    \frac{n'_{\rm DT}}{n'_{\rm BLR}} \approx \frac{\eta_{\rm DT}}{\eta_{\rm BLR}}\left(\frac{R_{\rm BLR}}{R_{\rm DT}} \right)^2\frac{\epsilon_{\rm BLR}}{\epsilon_{\rm DT}} \approx 0.1
    \label{eq:ratio-dens}
\end{equation}
where $\eta_{\rm DT} \sim \eta_{
\rm BLR}$ are the covering factors of the torus and the BLR, and $\epsilon_{\rm BLR}=2$~eV, $\epsilon_{\rm DT}=0.2$~eV, $R_{\rm BLR}=0.1$~pc, and $R_{\rm DT}=1$~pc. The ratio of the pion-production efficiencies can be then estimated as
\begin{equation}
    \frac{f_{\rm p\pi, DT}}{f_{\rm p\pi, BLR}}\approx \frac{n'_{\rm DT}}{n'_{\rm BLR}} \frac{R_{\rm DT}}{R_{\rm BLR}} \sim 1 ,
\end{equation}
where we assumed that the emitting region is located at the edge of the dusty torus and used Eq.~(\ref{eq:ratio-dens}). Therefore, for typical parameter values the pion production efficiencies are comparable~\citep[see also][]{Murase_2014, Oikonomou_2021}. However, for $\sigma\le 10$, the proton energy spectra are steep ($p>2$). As a result the increase in the proton energy threshold translates into a lower luminosity for the interacting protons, namely  $\varepsilon_{\rm p}L_{\varepsilon_{\rm p}}\large|_{\rm > th}\propto \varepsilon_{\rm p, th}^{-p+2}$. We can then estimate the ratio of the peak neutrino luminosities, for all other parameters fixed, as
\begin{equation}
    \frac{\varepsilon_{\nu} L_{\varepsilon_{\nu}}\large|_{\rm DT}}{\varepsilon_{\nu} L_{\varepsilon_{\nu}}\large|_{\large_{\rm BLR}}} 
    \approx \frac{f_{\rm p\pi, DT}}{f_{\rm p\pi, BLR}} \left(\frac{\epsilon_{\rm BLR}}{\epsilon_{\rm DT}}\right)^{-p+2} \approx 0.1~\frac{f_{\rm p\pi, DT}}{f_{\rm p\pi, BLR}},
\end{equation}
where the numerical value is computed for $p=3$. Therefore, the neutrino luminosity from pion production on the torus photons is expected to be lower than the one computed for the BLR photons.  

Radio observations of blazars indicate a correlation between the radio power and $\Gamma$. Based on this we introduced a power-law relation between the accretion rate and the jet Lorentz factor, $\dot{m}\propto \Gamma^s$ -- see Eq.~(\ref{eq:mdot}). This is another key point of our model, as it provides a way to associate low-$\sigma$ (high-$\Gamma$) jets with higher accretion rates and more luminous external photon fields -- see Eq.~(\ref{eq:Ld}). However, 
we did not choose the value of $s$ based on theoretical grounds. One therefore may ponder how our results would be affected if a different value of $s$ was adopted. The effects of $s$ on the (leptonic) photon SEDs were explored in detail in \cite{2021MNRAS.501.4092R}.  The authors showed that the exact value of $s$ changes the bolometric photon luminosity but not the spectral shape (see Fig.~A1 of their Appendix). This can be understood because the injection luminosity of particles is proportional to $L_{\rm j}$ which in turn depends on $\dot{m}$ -- see Eqs.~(\ref{eq:Lj}) and (\ref{eq:Lpart}). Meanwhile, the magnetic energy density is independent of $L_{\rm j}$, and the comoving energy density of external photons depends only on $\Gamma$. As a result, for an FSRQ-like source and a given pair of $(\mu, \sigma)$ values, a different value of $s$ would not change the ratio of $u'_{\rm B}/u'_{\rm BLR}$ (or the Compton ratio) but only the overall luminosity. Regarding the neutrino emission, any changes would be caused by changes in $L'_{\rm p}$ as long as the external photons would be the main targets for photopion interactions. For example, lower values of $s$ would lead to higher proton injection luminosities, and higher neutrino luminosities. However, in the FSRQ-like sources, the ratio $Y_{\nu \gamma}$ should remain unchanged, since $L_{\rm \gamma}\propto L_{\rm e}$. Similar trends are expected for the BL Lac-like sources in our model. The only difference is that a super-linear scaling relation of $L_{\nu+\bar{\nu}}$ on  $L_{\rm p}$ is expected because the target photon density will also depend on $L_{\rm e}$.

We simulated blazars covering a wide range of dimensionless accretion rates, $\dot{m} \sim 10^{-5}-1$, where $\dot{m}$ is defined in Eq.~(\ref{eq:2}). For simplicity, we assumed a linear scaling of the disc (and BLR) luminosity with $\dot{m}$. However, below a certain value for the accretion rate ($\sim 0.02$) the disc becomes less luminous than the prediction of Eq.~(\ref{eq:Ld}), because it becomes geometrically thick and radiatively inefficient \citep[e.g.][]{Sbarrato_2014}. Therefore, the BLR luminosity for the simulated blazars with $\sigma \ge 10$ should be lower than the one used here. This would not affect much the results for the strongly magnetized blazars ($\sigma=30, 50$), where the dominant seed photons for inverse Compton scattering and pion production are the synchrotron jet photons. For $\sigma=10$, however, a decrease in the BLR luminosity would decrease both the Compton luminosity and the peak neutrino luminosity.

\section{Conclusion}\label{sec:conclusion} 
We have presented a simple, but physically motivated, radiation model for baryon-loaded blazar jets. According to this, primary electrons and protons are accelerated to relativistic energies via magnetic reconnection in parts of the jet where the plasma magnetization is $\sigma\ge 1$. The blazar SED is produced by synchrotron and inverse Compton radiation of primary electrons. Electromagnetic emission produced  directly or indirectly by relativistic protons is in most cases subdominant. In our model, low-luminosity blazars ($L_{\gamma} \lesssim 10^{45}$~erg s$^{-1}$) are associated with less powerful and slower jets ($\Gamma \lesssim 5$) with higher magnetizations ($\sigma > 10$) in the jet location where energy dissipation takes place. Their broadband photon spectra resemble those of HSP BL Lac objects, and the expected neutrino luminosity is $L_{\nu+\bar{\nu}}\sim (0.3-1) \,  L_{\gamma}$. On the other end, high-luminosity blazars ($L_{\gamma} \gg 10^{45}$~erg s$^{-1}$) are associated with more powerful faster jets ($\Gamma > 10$) and lower magnetizations ($\sigma \le 10$). Their broadband photon spectra resemble those of FSRQs, while they are expected to be dim neutrino sources with $L_{\nu+\bar{\nu}}\ll L_{\gamma}$. The implications of our model for the diffuse neutrino flux from the blazar population are worth investigating and will be the subject of a future publication. 
 
\section*{Acknowledgements}
We would like to thank Paolo Padovani for useful comments on the manuscript. MP acknowledges support from the MERAC Fondation through the project THRILL and from the Hellenic Foundation for Research and Innovation (H.F.R.I.) under the ``2nd call for H.F.R.I. Research Projects to support Faculty members and Researchers'' through the grant number 3013 (UNTRAPHOB). DG acknowledges support from the Fermi Cycle 14 Guest Investigator Program 80NSSC21K1951, 80NSSC21K1938, and the NSF AST-2107802 and AST-2107806 grants. This research made use of Astropy,\footnote{\url{http://www.astropy.org}} a community-developed core Python package for Astronomy \citep{2013A&A...558A..33A, 2018AJ....156..123A}.

\section*{Data Availability}
All numerical models presented in this paper were computed using a proprietary numerical code. They can be shared upon reasonable request to the authors.
\bibliographystyle{mnras}
\bibliography{bibliography} 
\appendix 
\section{Code input parameters}\label{sec:appA}
We present in Tables~\ref{tab:mu50}-\ref{tab:mu90} the code input parameters for all the $(\mu,\sigma)$ values considered in this study. We note that the code takes as input a dimensionless form of the particle injection luminosities, the so-called compactness, which is defined as
\begin{equation}
\ell_{\rm e(p)}=\frac{L'_{\rm e(p)} \sigma_{\rm T}}{4 \pi R'_{\rm b} m_{\rm e(p)} c^{3}},
\label{eq:lelp}
\end{equation}
where $L'_{\rm e(p)}$ are given by Eq.~(\ref{eq:Lpart}). Similarly, the comoving energy density of the BLR photons is expressed through its compactness as 
\begin{equation}
    \ell_{\rm BLR}=\frac{u'_{\rm BLR} \sigma_{\rm T} R'_{\rm b}}{m_{\rm e} c^{2}}, 
    \label{eq:lext}
\end{equation} 
where $u'_{\rm BLR}$ is given by Eq.~(\ref{eq:uBLR}).

\begin{table*}
\centering 
\caption{Code input parameters used for the runs with $\mu = 50$.}
\begin{tabular}{c||ccccc}
\hline
\multicolumn{6}{c}{$\mu=50$} \\
\hline
$\sigma$     &  1 & 3 & 10 & 30 & 48.9\\
$\Gamma$     &  25 & 12.5 & 4.55 & 1.61 & 1.002 \\
$\dot{m}$    & 0.24 & 0.03 & $1.4 \times 10^{-3}$&  $6.6 \times 10^{-5}$ &  $1.6 \times 10^{-5}$ \\
$\delta$     & 28.38 & 20.98 & 8.77 & 2.87 & 1.07 \\
$R'_{\rm b}$~(cm) &  $2 \times 10^{16}$ & $1.4 \times 10^{16}$ &  $8.5 \times 10^{15}$ & $5.1 \times 10^{15}$ & $4 \times 10^{15}$  \\ 
$B'$~(G) &  8.6 & 10.6 & 11.75 & 13.52 & 47.93\\
$p$ & 3 & 2.5 & 2.2 & 1.5 & 1.2 \\
$\gamma'_{\rm min, e}$ & $10^{2.4}$ & $10^{2.7}$ & $10^{2.9}$ &  $10^3$ & $10^3$ \\
$\gamma'_{\rm min, p}$ & $10^{0.1}$ & $10^{0.1}$ & $10^{0.1}$ & $10^{0.1}$ & $10^{0.1}$\\
$\gamma'_{\rm max, e}$ &  $10^{6.1}$ & $10^{6.01}$ &  $10^{6.0}$ & $10^{6.0}$ & 
$10^{5.7}$ \\
$\gamma'_{\rm max, p}$ & $10^{8.6}$ & $10^{8.7}$ & $10^{8.6}$ & $10^{8.8}$ & $10^{8.6}$\\
$\ell_{\rm e}$ & $4 \times 10^{-3}$ & $4.3 \times 10^{-3}$ & $ 3.2 \times 10^{-3}$ &  $2.5 \times 10^{-3}$ &  $2.5 \times 10^{-2}$ \\
$\ell_{\rm p}$ & $2.2 \times 10^{-6}$ & $2.3 \times 10^{-6}$ & $1.7 \times 10^{-6}$ &  $1.4 \times 10^{-6}$ &  $1.3 \times 10^{-5}$ \\
$T'_{\rm BLR}$~(K) &  $2.2 \times 10^5$ & $1.1 \times 10^5$ &  $3.9 \times 10^4$ & $1.4 \times 10^4$ & $8.6 \times 10^3$ \\
$\ell_{\rm BLR}$ &  0.36  & $6.3 \times 10^{-2}$ & $5 \times 10^{-3}$ & 
 $3.4 \times 10^{-4}$ &  $8.7 \times 10^{-5}$ \\
 \hline
\end{tabular}
\label{tab:mu50}
\end{table*}

\begin{table*}
\centering 
\caption{Code input parameters used for the runs with $\mu = 70$.}
\begin{tabular}{c||ccccc}
\hline
\multicolumn{6}{c}{$\mu=70$} \\
\hline
$\sigma$     &  1 & 3 & 10 & 30 & 50\\
$\Gamma$     &  35 & 17.5 &  6.36 & 2.26 & 1.4 \\
$\dot{m}$    &  0.67 & 0.083 &  $4 \times 10^{-3}$ &  $1.8 \times 10^{-4}$ &  $4.3 \times 10^{-5}$ \\
$\delta$     & 28.09 & 25.49 & 12.06 & 4.26 & 2.31\\
$R'_{\rm b}$~(cm) & $2.4 \times 10^{16}$ &  $1.7 \times 10^{16}$ &  $1.1 \times 10^{15}$ &  $6 \times 10^{15}$ &  $4.7 \times 10^{15}$  \\ 
$B'$~(G) & 8.6 &  10.6 & 11.7 & 12.6 &  14.4 \\
$p$ & 3 & 2.5 & 2.2 & 1.5 & 1.2 \\
$\gamma'_{\rm min, e}$ & $10^{2.4}$ & $10^{2.7}$ & $10^{2.9}$ &  $10^3$ & $10^3$ \\
$\gamma'_{\rm min, p}$ & $10^{0.1}$ & $10^{0.1}$ & $10^{0.1}$ & $10^{0.1}$ & $10^{0.1}$\\
$\gamma'_{\rm max, e}$ &  $10^{6.1}$ & $10^{6.1}$ &  $10^{6}$ & $10^{6}$ & 
$10^{6}$ \\
$\gamma'_{\rm max, p}$ & $10^{8.5}$ & $10^{8.7}$ & $10^{8.7}$ & $10^{8.3}$ & $10^{8.9}$\\
$\ell_{\rm e}$ &  $4.7 \times 10^{-3}$ &  $5 \times 10^{-3}$ & $3.7 \times 10^{-3}$ &  $2.9 \times 10^{-3}$ &  $2.6 \times 10^{-3}$ \\
$\ell_{\rm p}$ &  $2.6 \times 10^{-6}$ &  $2.8 \times 10^{-6}$ & $2 \times 10^{-6}$ &  $1.4 \times 10^{-6}$ &  $1.4 \times 10^{-6}$ \\
$T'_{\rm BLR}$~(K) &  $3 \times 10^5$ &  $1.5 \times 10^5$ &  $5.5 \times 10^4$ & $1.9 \times 10^4$ &  $1.2 \times 10^4$  \\
$\ell_{\rm BLR}$ &  0.83 &  0.15 &  $1.2 \times 10^{-2}$ & 
 $8.4 \times 10^{-4}$ &  $2.3 \times 10^{-4}$ \\
 \hline
\end{tabular}
\label{tab:mu70}
\end{table*}

\begin{table*}
\centering 
\caption{Code input parameters used for the runs with $\mu = 90$.}
\begin{tabular}{c||ccccc}
\hline
\multicolumn{6}{c}{$\mu=90$} \\
\hline
$\sigma$     &  1 & 3 & 10 & 30 & 50\\
$\Gamma$     &  45 & 22.5 &  8.18 & 2.9 & 1.8 \\
$\dot{m}$    &  1.42 & 0.18 & $8.6 \times 10^{-3}$ & $3.8 \times 10^{-4}$ & $9.2 \times 10^{-5}$ \\
$\delta$     & 25.96 & 27.83 & 15.09 & 5.58 & 3.21\\
$R'_{\rm b}$~(cm) &  $2.7 \times 10^{16}$& $1.9 \times 10^{16}$ & $1.1 \times 10^{16}$ & $6.8 \times 10^{15}$ &  $5.4 \times 10^{15}$  \\ 
$B'$~(G) & 8.6 &  10.6 &  11.8 & 12.4 &  13.2 \\
$p$ & 3 & 2.5 & 2.2 & 1.5 & 1.2 \\
$\gamma'_{\rm min, e}$ & $10^{2.4}$ & $10^{2.7}$ & $10^{2.9}$ &  $10^3$ & $10^3$ \\
$\gamma'_{\rm min, p}$ & $10^{0.1}$ & $10^{0.1}$ & $10^{0.1}$ & $10^{0.1}$ & $10^{0.1}$\\
$\gamma'_{\rm max, e}$ &  $10^{6.1}$ & $10^{6.1}$ &  $10^{6}$ & $10^{6}$ & $10^{6}$ \\
$\gamma'_{\rm max, p}$ & $10^{8.4}$ & $10^{8.7}$ & $10^{8.7}$ & $10^{8.8}$ & $10^9$\\
$\ell_{\rm e}$ &  $5.4 \times 10^{-3}$ &  $5.6 \times 10^{-3}$ &  $4.2 \times 10^{-3}$ &  $2.8 \times 10^{-3}$ & $2.5 \times 10^{-3}$   \\
$\ell_{\rm p}$ & $2.9 \times 10^{-6}$ & $3.1 \times 10^{-6}$ & $2.3 \times 10^{-6}$ &  $1.5 \times 10^{-6}$ & $1.4 \times 10^{-6}$ \\
$T'_{\rm BLR}$~(K) & $3.9 \times 10^5$ &  $1.9 \times 10^5$ & $7 \times 10^4$ & $2.5 \times 10^4$ &  $1.6 \times 10^4$ \\
$\ell_{\rm BLR}$ &  1.55 &  0.27 &  $2.2 \times 10^{-2}$ & $1.6 \times 10^{-3}$ &  $4.6 \times 10^{-4}$ \\
 \hline
\end{tabular}
\label{tab:mu90}
\end{table*}

\section{Semi-analytical calculation of neutrino spectra}\label{sec:appB}
We can estimate in a semi-analytical manner the differential all-flavour neutrino luminosity as
\begin{equation} 
\varepsilon_{\nu} L_{\varepsilon_{\nu + \bar{\nu}}} \approx \frac{3}{8} f_{\rm p\pi} \delta^4 \varepsilon'_{\rm p} L'_{\varepsilon'_{\rm p}}
\label{eq:Lv}
\end{equation}
where $L'_{\varepsilon'_{\rm p}} \equiv {\rm d}L'_{\rm p}/{\rm d}\varepsilon'_{\rm p}$  is the differential proton luminosity in the comoving frame, $\delta$ is the Doppler factor of the emitting region, and $f_{\rm p\pi}$ is the photopion production efficiency.  The latter is defined as $f_{\rm p\pi} = 1/(1+t^\prime_{\rm p \pi}/t'_{\rm dyn})$, where $t'_{\rm dyn}=R'_{\rm b}/c$ and $t'_{\rm p \pi}$ is the proton energy loss timescale due to photopion production. 

For an isotropic radiation field the inverse of the energy loss timescale for a proton with Lorentz factor $\gamma_{\rm p}^{\prime}$ is calculated as \citep{Stecker_1968, Begelman_1990}
\begin{equation}
    t_{\rm p\pi}^{\prime-1} \left(\gamma_{\rm p}^{\prime}\right)= \frac{c}{2\gamma^{\prime 2}_{\rm p}} \int_{\bar{\epsilon}_{\rm th}}^\infty {\rm d}\bar{\epsilon}  \, \kappa_{\rm p \pi}(\bar{\epsilon})\sigma_{\rm p\pi}(\bar{\epsilon}) \bar{\epsilon}   \int_{\bar{\epsilon}_{{\rm th}} / 2 \gamma_{\rm p}^{\prime}}^{\infty} {\rm d} \varepsilon^{\prime} \frac{n_{\rm ph}^{\prime}\left(\varepsilon^{\prime}\right)}{\varepsilon^{\prime 2}}
\end{equation}
where $\sigma_{\rm p\pi}$ and $\kappa_{\rm p\pi}$ are the cross-section and proton inelasticity, respectively, $\bar{\epsilon}$ is the interaction energy (or the photon energy in the proton rest frame), and $n_{\rm ph}^{\prime}\left(\varepsilon^{\prime}\right)$  is the differential photon number density in
the comoving frame of the emission region. 

The target photon field has two contributions, from the BLR and the jet. Given that the non-thermal photons from the jet are produced by primary electrons, we can use the steady-state non-thermal photon spectra computed numerically with the {\sc athe$\nu$a} code as input in the above integral. We also adopt the total photopion production cross section from \cite{2019JCAP...11..007M} and we numerically compute the double integral (assuming $\kappa_{\rm p \pi}=0.2$) for various magnetizations and $\mu=50$.

\begin{figure}
    \centering
    \includegraphics[width=0.47\textwidth]{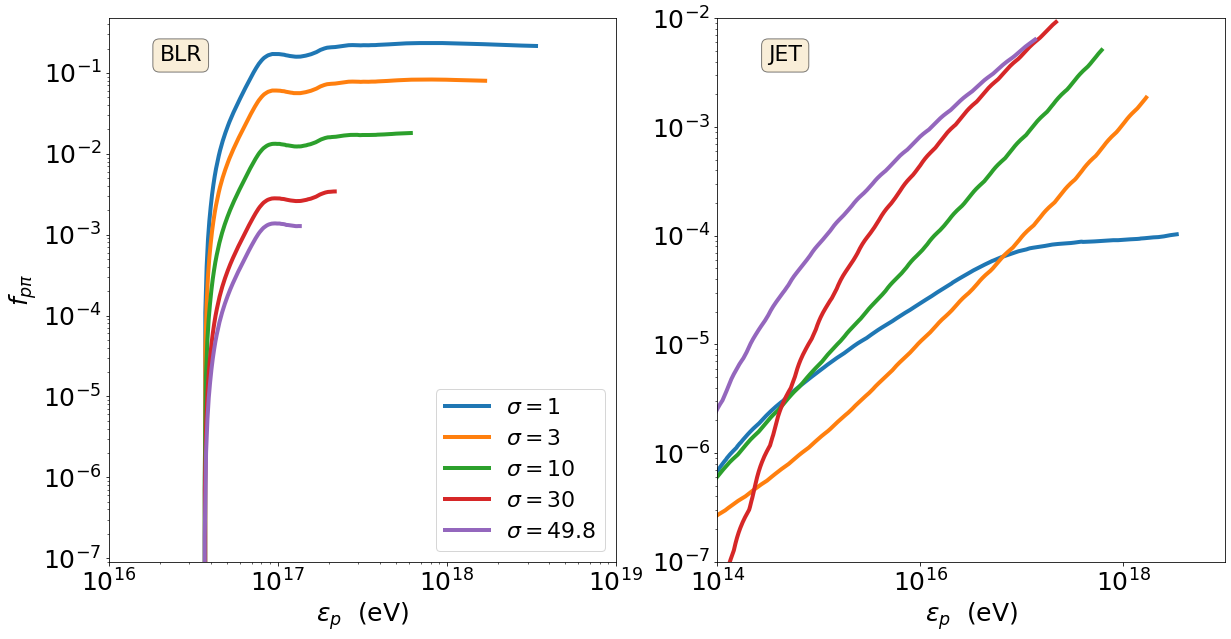}
    \caption{Efficiency of photopion production interactions with BLR photons (left) and jet photons (right) plotted as a function of the proton energy (in the observer's) frame for different values of $\sigma$ (see inset legend) and $\mu=50$. }
    \label{fig:fpg}
\end{figure}

Our results for $f_{\rm p \pi}$ are shown in Fig.~\ref{fig:fpg}. The efficiency computed using only BLR or jet photons is displayed separately in the two panels. Interactions with BLR have higher efficiency across all proton energies for $\sigma=1-10$, while interactions with jet photons are the main channel for neutrino production for $\sigma=30,50$. 

But in order to fully understand the dependence of the neutrino luminosity on $\sigma$, we need to check how the proton power, at energies that are relevant for neutrino production, scales with $\sigma$. In the absence of cooling, the differential proton luminosity can be written as
\begin{equation}
   L'_{\varepsilon'_{\rm p}}= V'Q'_{\rm 0, p} \left(\frac{\varepsilon'_{\rm p}}{m_{\rm p} c^2} \right)^{-p+1}
\end{equation}
where $V'Q'_{\rm 0, p}$ is given by Eq.~(\ref{eq:Qo}). The comoving proton luminosity for interactions with BLR photons above the threshold then reads 
\begin{equation}\label{eq:4.thr}
   L^{\prime}_{\rm p}\large|_{>\rm th} = L'_{\rm p}  \frac{\varepsilon^{'2-p}_{\rm max, p}  - \varepsilon^{'2-p}_{\rm th, p}}{\varepsilon^{'2-p}_{\rm max, p}  - \varepsilon^{'2-p}_{\rm min, p}}
\end{equation}
where the proton energy threshold for interactions with BLR photons of energy 2~eV is $\varepsilon'_{\rm th,p} \simeq 3\times 10^{16}~\rm eV/\Gamma$. Meanwhile, the threshold condition for interactions with jet photons is always satisfied; the lowest energy protons can interact with inverse Compton scattered (ICS) photons, but the efficiency of the interaction is very low (not explicitly shown in the figure) because of the low ICS photon number density.

\begin{figure}
    \centering
    \includegraphics[width=0.47\textwidth]{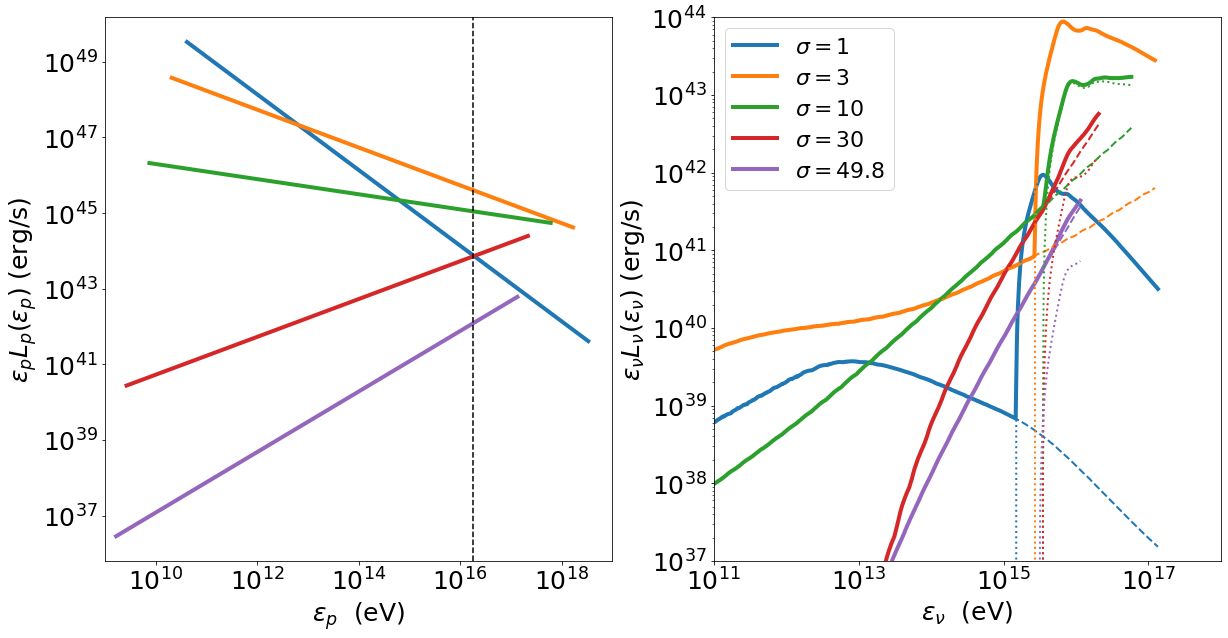}
    \caption{Left: differential proton luminosity (in the observer's frame) plotted against the proton energy for $\mu=50$ and various values of $\sigma$ (see inset legend). The vertical line marks the proton energy threshold with BLR photons of energy 2~eV (in the AGN rest frame). Right: spectrum of all-flavour neutrino luminosity (in the observer's frame) computed using Eq.~(\ref{eq:Lv}). Contributions from the BLR and jet photons are plotted with dotted and dashed lines, respectively.}
    \label{fig:Lp-Lv}
\end{figure}
In the left panel of Fig.~\ref{fig:Lp-Lv} we present the spectrum of proton luminosity (in the observer's frame) for different plasma magnetizations. The vertical line indicates the threshold energy for interactions with BLR photons of energy 2~eV. The integrated proton luminosity above that energy becomes maximum for $\sigma=3$. Given that this is about two orders of magnitude higher than $L_{\rm p}\large|_{>\rm th}$ for $\sigma=1$ and that the BLR photopion production efficiency for $\sigma=3$ is about 10 times lower than the one for $\sigma=1$ (see left panel in Fig.~\ref{fig:fpg}), the neutrino luminosity becomes also maximum for $\sigma=3$ -- see right panel of Fig.~\ref{fig:Lp-Lv}. Notice also that for $\sigma \le 10$, the neutrino spectra have two components, with the one peaking at highest energies resulting from interactions with BLR photons. For the high-$\sigma$ cases, however, the neutrino spectrum is dominated by interactions with the jet synchrotron photons. These results are in agreement with those obtained with the full numerical code (see e.g. Fig.~\ref{fig:seds-2}).

\section{Steady-state lepton distributions}\label{sec:appC}
\begin{figure*}
    \centering
    \includegraphics[width=\textwidth]{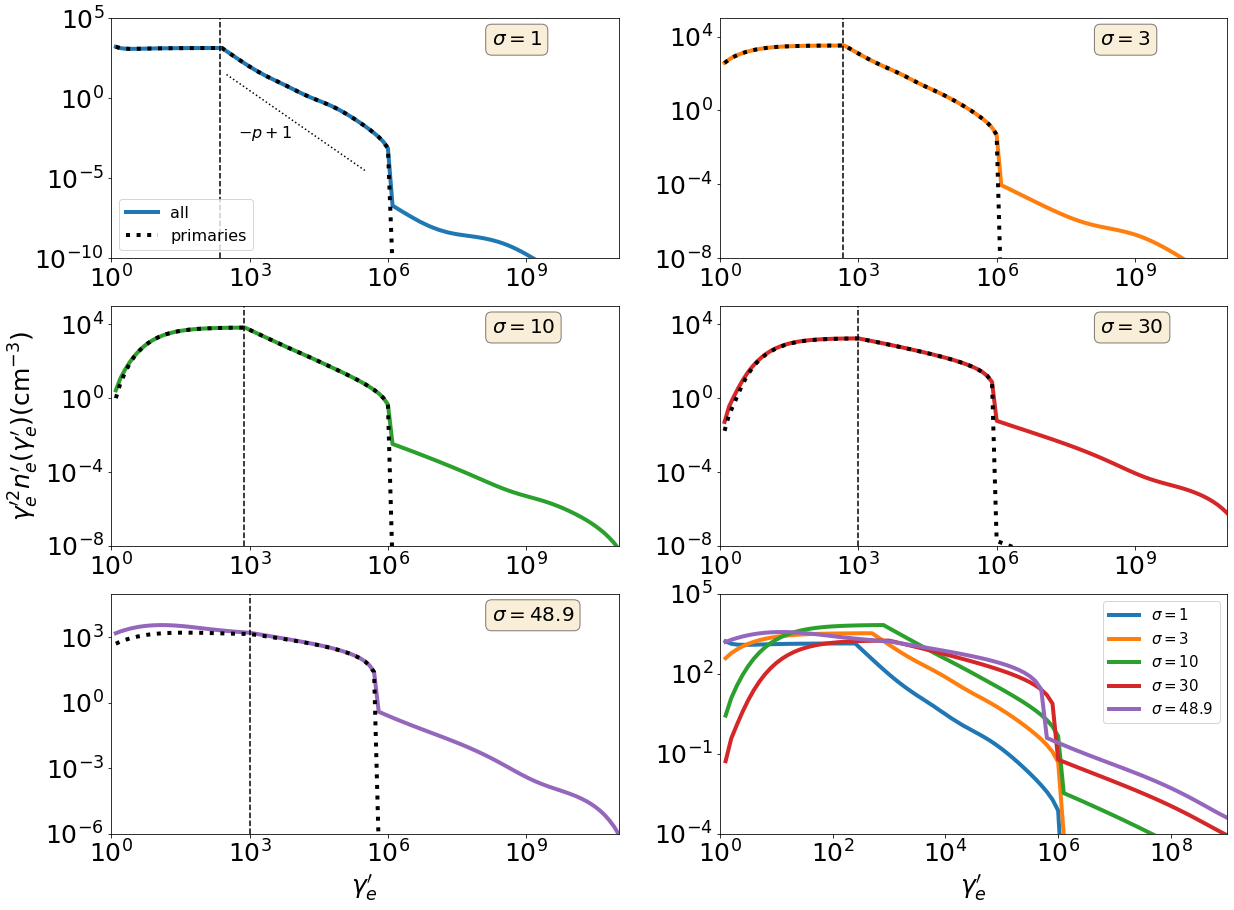}
    \caption{Steady-state lepton energy distributions (in the comoving frame) for $\mu=50$ and different values of $\sigma$. The primary contribution to the total spectrum (thick coloured line) is shown with a dashed line. The vertical dashed line marks $\gamma'_{\min, \rm e}$. The total lepton energy spectra for five values of $\sigma$ are are plotted for easier comparison in the lower right panel.}
    \label{fig:spec-elec}
\end{figure*}

Figure~\ref{fig:spec-elec} shows the steady-state differential density distributions of leptons, compensated by $\gamma_{\rm e}^{\prime 2}$, in the comoving frame for $\mu=50$ and different values of $\sigma$. Similar results are found for $\mu=70, 90$ and for this reason are not displayed. 

For each value of $\sigma$ we show separately the distribution of primary (i.e. accelerated leptons). 
For $\gamma'_{\rm e} > \gamma'_{\rm max, e}$, where no primaries are injected, we can clearly see the contribution of secondaries produced via photohadronic and $\gamma \gamma$ pair production processes. However, their emission is negligible to the total photon spectrum,  since these energetic leptons carry only a small amount of the available energy. Generally, the contribution of secondaries to the total distribution becomes progressively more important for higher magnetizations. This results also agrees with the trend found for $L_{\nu+\bar{\nu}}/L_{\gamma}$ with $\sigma$ (see Fig.~\ref{fig:Y-Lg}). Still, the secondary contribution to the energy range where primaries are injected is subdominant except for $\sigma=48.9$; in this case, secondaries that cool down to low energies produce an excess over the primary distribution.

In all steady-state spectra shown in the figure we see evidence of particle cooling below the minimum injection Lorentz factor (marked with a vertical line). In this regime, also known as \textit{fast cooling}, even electrons injected at $\gamma'_{\min, \rm e}$ can cool within one dynamical timescale, thus producing the low-energy extension with the characteristic slope of $-2$. In this case, the (primary) particle distribution can be expressed as
\begin{equation}
n'_{\rm e}\left(\gamma_{\rm e}^{\prime}\right) \propto \begin{cases}\gamma_{\rm e}^{\prime-2}, & \gamma_{\rm c, e}^{\prime}<\gamma_{\rm e}^{\prime} \le \gamma_{\min, \rm e}^{\prime} \\ \gamma_{\rm e}^{\prime-p-1}, & \gamma_{\min, \rm e }^{\prime}  < \gamma_{\rm e}^{\prime} \le \gamma^\prime_{\max, \rm e} \end{cases}
\end{equation}
where $p$ is the power-law slope at injection and $\gamma'_{\rm c, e}$ is the cooling Lorentz factor, which is defined as $t'_{\rm loss, e}=t'_{\rm dyn}=R'_{\rm b}/c$. Here, $t'_{\rm loss, e}(\gamma'_{\rm e})$ is the energy loss timescale of electrons due to synchrotron and inverse Compton scattering. The cooling is stronger for lower magnetizations where  $\gamma'_{\rm c, e}\approx 1$, while it becomes a bit weaker for $\sigma=10-30$,  where $\gamma'_{\rm e, c}\approx 0.1 \gamma_{\rm min, e}$. 

Close inspection of the spectrum for $\sigma=1$ shows a spectral break at $\gamma'_{\rm e} \approx 10^4$, with the spectrum becoming less steep than the prediction ($-p+1$, see dotted line). Keeping in mind that for low magnetizations electrons are predominantly cooling via inverse Compton scattering off the boosted BLR photons ($\epsilon'_{\rm BLR}=40 (\Gamma/20)~{\rm eV}$), this spectral break marks the transition from Thomson to Klein-Nishina cooling at $\gamma'_{\rm e} \approx 3 m_{\rm e} c^2/4\epsilon'_{\rm BLR}$. It is also reflected on the synchrotron spectrum (see left panel in Fig.~\ref{fig:seds-1}). This transition is also present in the spectra for $\sigma=3$, but less evident. We note that this spectral break was missed by \cite{2021MNRAS.501.4092R} where electron cooling only in the Thomson regime was considered.


\bsp	
\label{lastpage}
\end{document}